\documentclass{amsart}
\usepackage[utf8]{inputenc}
\usepackage{amsmath}
\usepackage{array, amsfonts}
\usepackage{amsthm}
\usepackage{amssymb}
\usepackage{enumerate}
\usepackage{lineno}
\usepackage[bookmarks=true]{hyperref}
\usepackage{enumitem}
\usepackage{mathrsfs}
\usepackage{stackengine}
\usepackage{bookmark}
\usepackage{color,soul}
\usepackage{todonotes}
\usepackage{graphicx}
\usepackage{subfigure}
\usepackage{balance}
\usepackage{textcomp}
\usepackage{xcolor}
\usepackage[foot]{amsaddr}

\definecolor{darkgreen}{rgb}{0,0.5,0}

\newcommand{\comment}[1]

\usepackage[numbers,sort&compress]{natbib}

\DeclareMathOperator*{\myDelta}{\Delta}

\title{Emergence and algorithmic information dynamics of systems and observers 
}

\author[Felipe S. Abrah\~{a}o]{Felipe S. Abrah\~{a}o}
\address[Felipe S. Abrah\~{a}o]{National Laboratory for Scientific Computing (LNCC), 25651-075, Petropolis, RJ, Brazil. Algorithmic Nature Group, Laboratoire de Recherche Scientifique (LABORES) for the Natural and Digital Sciences, 75005, Paris, France. }
\thanks{F. S. Abrah\~{a}o  acknowledges partial support from CNPq (301429/2021-9), INCT in Data Science – INCT-CiD (CNPq 465.560/2014-8), and FAPERJ (E-26/203.046/2017).}
\email[Felipe S. Abrah\~{a}o]{fsa@lncc.br}

\thanks{We thank  Itala M. L. D'Ottaviano, Leonardo L. de Carvalho, Virginia M. F. G. Chaitin, Michael Winter, Alyssa Adams, Hyunju Kim, Sara I. Walker, Paul Davies, and Mikhail Prokopenko for suggestions and rewarding discussions on related topics.}

\author[Hector Zenil]{Hector Zenil}  
\address[Hector Zenil]{ Oxford Immune Algorithmics, RG1 3EU, Reading, U.K.. Alan Turing Institute, British Library, 2QR, 96 Euston Rd, London NW1 2DB. Algorithmic Dynamics Lab, Unit of Computational Medicine, Department of Medicine Solna, Center for Molecular Medicine, Karolinska Institute, SE-171 77, Stockholm, Sweden. Algorithmic Nature Group, Laboratoire de Recherche Scientifique (LABORES) for the Natural and Digital Sciences, 75005, Paris, France.
}
\email[Hector Zenil]{hector.zenil@cs.ox.ac.uk}

\begin{document}

\maketitle

\newtheorem{observationprinciple}{\bf Observation principle}
\newtheorem{idealobservationprinciple}{\bf Ideal observation principle}
\newtheorem{definition}{\bf Definition}[section]
\newtheorem{lemma}{\bf Lemma}[section]
\newtheorem{theorem}{\bf Theorem}[section]
\newtheorem{condition}{\bf Condition}[section]
\newtheorem{corollary}{\bf Corollary}[section]

%
%
%
%
%

\begin{abstract}
Previous work has shown that perturbation analysis in software space can produce candidate computable generative models and uncover possible causal properties from the finite description of an object or system quantifying the algorithmic contribution of each of its elements relative to the whole.
One of the challenges for defining emergence is that one observer's prior knowledge may cause a phenomenon to present itself to such observer as emergent while for another as reducible.
By formalising the act of observing as mutual perturbations between dynamical systems, 
we demonstrate that emergence of algorithmic information do depend on the observer's formal knowledge, while robust to other subjective factors, particularly: the choice of the programming language and the measurement method; errors or distortions during the information acquisition; and the informational cost of processing.
This is called observer-dependent emergence (ODE).
In addition, we demonstrate that the unbounded and fast increase of emergent algorithmic information implies asymptotically observer-independent emergence (AOIE). 
Unlike ODE, AOIE is a type of emergence for which emergent phenomena will remain considered to be emergent for every formal theory that any observer might devise.
We demonstrate the existence of an evolutionary model that displays the diachronic variant of AOIE and a network model that displays the holistic variant of AOIE.
Our results show that, restricted to the context of finite discrete deterministic dynamical systems, computable systems, and irreducible information content measures, AOIE is the strongest form of emergence that formal theories can attain. 
\end{abstract}



\pagebreak
\section{Introduction}

Perturbation (or intervention) analyses of changes in the algorithmic information necessary for computably constructing an object enable the investigation of the underlying causal effectiveness of its parts (or elements) \cite{Zenil2020a,Zenil2019a} as well as the solution of the inverse problem of finding the best generative model \cite{Zenil2019}.
This approach, elaborated within
the framework of algorithmic information dynamics \cite{Zenil2020}, is based on the (the expected) universal optimality of algorithmic probability \cite{Chaitin1987,Li1997,Downey2010,Calude2002} and stems from the demonstrated high convergence rate of computable generative models to the algorithmic probability \cite{Zenil2020b}.
Working within this framework, the present article tackles the problem of quantifying the emergence of algorithmic information in discrete deterministic dynamical systems and computable systems.

The challenge of formalising the notion of \emph{emergence} usually inheres in the definition of what the term "reducibility" ("derivable" or "predictable") means when one says that a macro-level phenomenon is not reducible to its micro-level parts or to its initial conditions.
In order to eliminate the possibility of one observer classifying a phenomenon as emergent while another  classifies it as reducible to its isolated parts (to the parts at a
smaller scale or to the initial conditions), one approach is to define emergence as a property relative to the micro-level parts or to the initial conditions \cite{Bedau1997,Adams2017,Hernandez-Orozco2018,Abrahao2019}.
The mathematical and empirical problem is to guarantee that
such a dependency on the observer cannot occur even when formalising emergence as a relative property \cite{Chalmers2008}. 
In this article, to tackle this problem in the context of finite discrete deterministic dynamical systems (or computable systems in general), the act of observing is formally defined as an interaction in which the system being observed perturbs the observer while the observer perturbs the system being observed, where the observer is a particular type of system that can compute functions and is equipped with a formal theory. 
Hence, we show that mathematical measures of emergent behaviour as a relative property do depend on the formal theories that the observer brings to bear.

Although being dependent on the observer's formal knowledge, we show that the emergence of algorithmic information is robust in the face of variations of the arbitrarily chosen method of measuring irreducible information content, errors (or distortions) in the very act of observing, and variations of the algorithmic-informational cost of processing the information gathered from the observed system in accordance with the observer's formal knowledge.
In other words, all the subjective factors of language, measurement, information acquisition, and processing are embedded into the definition of emergence of algorithmic information in such a way that we demonstrate that the formal theory (which the very observer has brought to bear) is the only subjective characteristic of the observer that can determine whether or not the future behaviour of the observed system will appear emergent.
This kind of emergence is called \emph{observer-dependent emergence} (ODE).

Furthermore, we show that systems that display a sufficiently fast increase of emergent algorithmic information overcome such a dependency on the observer.
In other words, there are systems whose behaviour eventually begins to display ODE for any observer.
Although there might be an observer that can explain or predict a finite-length state space trajectory of an observed system, the sufficiently fast increase of emergent algorithmic information guarantees that this will eventually cease to happen.
In this case, the emergence of algorithmic information is guaranteed to be independent of any observer, but only at the asymptotic limit.
This kind of emergence is called \emph{asymptotically observer-independent emergence} (AOIE).
The definition of AOIE inherits from ODE the robustness to the subjective factors of language, measurement, information acquisition, and processing.
However, unlike ODE, one aspect of AOIE that is remarkable is the fact that this is a type of emergence for which emergent phenomena will remain considered to be emergent for every formal theory that one might devise.

To achieve its results, this article introduces new definitions, lemmas, and theorems.
We also compare these mathematical properties with previous models and definitions in the literature that deal with emergence in discrete deterministic dynamical systems and computable systems and with definitions of weak and strong emergence.
We present an evolutionary model that displays the temporal (or diachronic) variant of AOIE and a model for networked systems that displays the holistic variant of AOIE.
In particular, the latter model displays expected downward causation.
Our results demonstrate that, restricted to the context of finite discrete deterministic dynamical systems, computable systems, and irreducible information content measures, AOIE is the \emph{strongest} form of emergence that a formal theoretical approach can grasp.
Future research is necessary for investigating whether or not the results in the present article can be extended to other physical, chemical, or biological systems and other complexity measures.

In Section~\ref{sectionAICandSystems}, we study how algorithmic information content can be quantified in finite discrete deterministic dynamical systems and computable systems. 
In Section~\ref{sectionObserversystems}, we introduce \emph{algorithmic perturbations}, \emph{formal observer systems}, and the minimum requirements (stated as the \emph{observation principle}) for the observation to take place.
In Section~\ref{sectionWeakemergentalgorithmicinformation}, we introduce ODE and analyse two previous works.
In Section~\ref{sectionAOIE}, we introduce AOIE, demonstrate that two previous mathematical models in the literature display AOIE, and analyse these models, comparing them to other approaches to weak and strong emergence.
Section~\ref{sectionConclusion} concludes the article.

%
%
%

\section{Algorithmic information content of objects and dynamical systems}\label{sectionAICandSystems}

For systems composed of (or defined by) stochastic processes, emergence has been studied in terms of statistical methods (for example, those based on entropy) and related complexity measures \cite{Prokopenko2009,Fernandez2014}.
If an independent and identically distributed (i.i.d.) stochastic process $ \left\{ \mathcal{ X }_i  \right\} $, where $ \mathcal{ X }_i $ is a random variable, produces sequences of (finite) states, one knows from the noiseless source coding theorem that $ n \, H\left( \mathcal{ X } \right) $ gives a lower bound for the minimum expected number of bits to encode a sufficiently long sequence generated by this stochastic process \cite{Cover2005}, where $ H\left( \mathcal{ X } \right) $ is the entropy and $ n $ is the length of the sequence.
In this context, due to such a minimality displayed by the entropy value in pure stochastic processes, emergence of novel irreducible information can, for example, be understood as an entropy increase, as proposed in \cite{Fernandez2014}.
On the other hand, when emergence is interpreted as the appearance of a macro-level property that has greater efficiency of prediction than that of the micro-level states from which the macro-states derive, emergence has been proposed to be measured by employing a ratio between excess entropy and statistical complexity \cite{Shalizi2001,Prokopenko2009}.
In the context of multivariate stochastic processes, causal emergence and downward causation have been proposed to be measured by employing variants of the unique information, which are based on the partial information decomposition \cite{Lizier2018} and integrated information decomposition \cite{Mediano2019}.

However, in the context of deterministic processes,
statistics faces insuperable obstacles when trying to quantify irreducible information content \cite{Zenil2020b}.
Being one of the well-known results in algorithmic information theory (see Section~\ref{sectionAICofOs}),
any resource-bounded computational procedure that tries to quantify the amount of irreducible information content in a single encoded object returns distorted values in general.
Although the entropy of its contiguous blocks of length $ m $ is maximal, this distortion is, for example, seen in Borel-normal sequences of length $ n $ that are in fact computable (and therefore logarithmically compressible) \cite{Becher2002}, where $ m \ll n $.
In the context of networks and graphs, there are also highly compressible graphs in which the degree-sequence entropy is maximal \cite{Zenil2017}.
Thus, if one is interested in measuring irreducible information content (or measuring the emergence of new irreducible information) in deterministic systems, which are free of stochasticity, employing any fixed and computable measure based only on finding and exploiting statistical patterns (in order to approximate the most compressed form that computes the system's behaviour) will exhibit limitations and face these obstacles in general.
The main limitation stems from the fact that most computable patterns are not periodic, the kind of regularity that a statistical approach would be able to characterise.  
Computable but non-periodic patterns will tend to have high statistical complexity (e.g. Shannon entropy with no access to the underlying probability distribution) but low algorithmic complexity, meaning that a statistical approach would assign them a random nature that, for all mechanistic and cause-and-effect purposes, should not.

The present article only addresses discrete \emph{deterministic} dynamical systems (or \emph{computable} systems in general). 
As we will show in the next Section~\ref{sectionAICofOs}, algorithmic-information-based approximation methods to the size of the irreducible information content are proved to be accurate in the asymptotic limit when the computational resources are unbounded.
In addition, due to the property of always existing ``room for improvement'' in resource-bounded compression algorithms, empirical applications of the theoretical results presented in this article are agnostic with respect to the chosen compression algorithm.

Future research is necessary for investigating in which conditions the theoretical results in the present article can be extended to stochastic processes and other complexity measures.

The introduction of perturbation (or intervention) analysis, in the context of algorithmic information content, enables the investigation of the underlying computable causal effectiveness of its parts (or elements) \cite{Zenil2020a,Zenil2019a} and offers a solution to the inverse problem of finding the best generative model \cite{Zenil2019}.  
Generative model in the context of algorithmic information means a step-by-step computable model (which in turn means being able to be carried out by a Turing machine) that generates the object, data sample, or system to be analysed.  
Such an introduction of perturbation analysis led to the introduction of algorithmic information dynamics (AID) \cite{Zenil2020}, based on the (expected) universal optimality of algorithmic probability (see Section~\ref{sectionAICofOs}) and stems from the demonstrated high convergence rate of computable generative models to their algorithmic probability \cite{Zenil2020b}. 
This is a rate stable to radical changes to the model of computation, that produces a stable distribution in particular for low complexity and thus high algorithmic frequency (probability) objects.  
(Note that the ultimate convergence is guaranteed by the invariance theorem \cite{Li1997,Downey2010,Calude2002,Chaitin1987}).
Working within this framework, the present article tackles the problem of quantifying the emergence of algorithmic information in discrete deterministic dynamical systems and computable systems.
In this regard, we show in this article that one of the paradigm shifts brought about by algorithmic information dynamics vis-\`{a}-vis previous methods based on computability and information theory is that perturbation analysis guarantees that our results hold, even if we allow the very act of observing to substantially change (or introduce ``noise'' into) the observed system's behaviour.

\subsection{Encoded objects}\label{sectionAICofOs}

Let $ \left| x \right| $ denote the length of a string $ x \in  \{ 0 , 1 \}^* $.
A universal programming language is said to be prefix-free if no string of this language is a prefix of other string in the same language.
Let $ \mathbf{U} $ denote a universal Turing machine that runs on a prefix-free universal programming language.
Let $ \mathbf{U}(x) $ denote the output of the universal (prefix) Turing machine $\mathbf{U}$ when $x$ is given as input in its tape.
Let $ \left< \, \cdot \, , \, \cdot \, \right> $ denote an arbitrary recursive bijective pairing function \cite{Li1997,Downey2010} so that the bit string $ \left< \, \cdot \, , \, \cdot \, \right> $ encodes the pair $ \left( x , y \right) $, where $ x , y \in \mathbb{N} $.
This notation can be recursively extended to $ \left< \cdot \, , \, \dots \, , \, \cdot   \right> $ in order to represent the encoding of $n$-tuples.
The Big-\textbf{O} notation $f(x)=\mathbf{O}( g(x) )$ denotes the usual \emph{weak} asymptotic dominance when function $f$ is asymptotically upper bounded by function $g$.
The (prefix) \emph{algorithmic complexity}
$ \mathbf{K}\left( x \right) $ is the length of the shortest prefix-free (or self-delimiting) program (which is denoted by $ x^* $) that outputs the string $ x $ in a universal prefix Turing machine $ \mathbf{U} $, i.e., $ \mathbf{U}\left( x^* \right) = x $ and the length $ \left| x^* \right| = \mathbf{K}\left( x \right) $ is minimum.
Other variants of algorithmic complexity in AIT include:
the \emph{conditional} prefix algorithmic complexity of a binary string $ z $ given a binary string $ w $, denoted by $ \mathbf{K}( z \, \vert w ) $, which is the length of the shortest program $ z_w^*$ such that $ \mathbf{U}( \left< w , z_w^* \right> ) = z $;
$ \mathbf{ I }_{ \mathbf{ K } }( w : z ) = \mathbf{K}(z) - \mathbf{K}( z \, | w ) $, which is the \textbf{K}-complexity of information in $ w $ about $ z $ and it quantifies the amount of irreducible information in $ w $ about $ z $; and
the \emph{mutual algorithmic information}
$  \mathbf{ I_A }( w \, ; z ) = \mathbf{K}(z) - \mathbf{K}( z \, | w^* ) $ between the arbitrary strings $w$ and $z$, which quantifies the amount of irreducible information in $ w $ about $ z $, and vice-versa.

When dealing with other kind of objects that are not strings, a mathematical object is said to be \emph{encoded} if it is univocally represented by structured data so that there is an algorithm which can always recover or extract the original object from the structured data.
A trivial example is encoding a (directed) graph as an indexed list of the characteristic function of the edges in the form $ \left( \left( v_1 , v_2 \right) , z_1 \right) \cdots  \left( \left( v_i , v_j \right) , z_k \right) \cdots  \left( \left( v_{ n } , v_{ n - 1 } \right) , z_{ n^2 - n } \right) $, where $ n $ is the number of vertices, $ n^2 - n $ is the total number of possible (directed) edges, $ 1 \leq i \leq n $, $ 1 \leq j \leq n $, and  $ 1 \leq k \leq n^2 - n $.
This way, one can equivalently define $ \mathbf{K}\left( x \right) $ (and all of the other above variants) when $ x $ is an encoded object instead of a string.

Algorithmic complexity is a \emph{invariant} measure of irreducible information content because, for any arbitrarily chosen encoding method or universal prefix-free programming language, the value can only vary by a constant that does not depend on the object \cite{Li1997,Downey2010,Calude2002,Chaitin1987}.
It is \emph{minimal} because, for any arbitrarily chosen computably enumerable semimeasure $ \mu\left( \cdot \right) $ of the infinite discrete space of all encoded finite objects, 
the value $ - \log\left( \mu\left( x \right) \right)  $ can only be smaller than the algorithmic complexity $ \mathbf{K}\left( x \right) $ up to an object-independent constant  \cite{Li1997,Downey2010}. 
This minimality can be easily verified as a consequence of the fact that $ \mathbf{K}\left( x \right) $ and the universal probability of occurring a computably constructible object $ x $ are inherently associated by the \emph{algorithmic coding theorem} in algorithmic information theory (AIT) \cite{Li1997,Downey2010,Calude2002,Chaitin1987}, which states that
\begin{equation}
	\mathbf{K}\left( x \right) = - \log\left( \sum\limits_{ \mathbf{U}\left( p \right) 
		= x } \frac{ 1 }{ 2^{ \left| p \right| } } \right) \pm \mathbf{O}( 1 )
	= - \log\left( \mathbf{m}\left( x \right) \right) \pm \mathbf{O}( 1 )
	\text{ ,}
\end{equation}
where:
$ \mathbf{m}\left( \cdot \right) $ is a maximal computably enumerable semimeasure; 
and $ \sum\limits_{ \mathbf{U}\left( p \right) 
	= x } 2^{ - \left| p \right| }  $
is called the universal a priori probability of $ x $.
The \emph{universal a priori probability} of $ x $ can be understood as the probability of randomly generating (by an i.i.d. stochastic process) a prefix-free (or self-delimiting) program that outputs $ x $. 
A computably enumerable semimeasure $ \mathbf{m}\left( \cdot \right) $ is said to be \emph{maximal} if, for any other computably enumerable semimeasure $ \mu\left( \cdot \right) $ with domain defined for possible encoded objects, where $ \sum\limits_{ x \in \left\{ 0 , 1 \right\}^* } \mu\left( x \right) \leq 1 $, there is a constant $ C > 0 $ such that, for every encoded object $ x $,
$ \mathbf{m}\left( x \right) \geq C \, \mu\left( x \right)\text{ .} $
In particular, we know from AIT that any \emph{computable} measure (or semimeasure) $ \mu'\left( \cdot \right) $ of the infinite discrete space of all encoded finite objects loses the property of being maximal \cite{Li1997}, unlike the computably enumerable semimeasure $ \mathbf{m}\left( \cdot \right) $ which is maximal.
Also note that the algorithmic coding theorem applies analogously to the conditional algorithmic complexity $ \mathbf{K}( z \, | w ) $.

Other remarkable properties of algorithmic complexity is that in general any compression algorithm can only approximate the value of $ \mathbf{K}\left( x \right) $ from above in such a way that it is not in general decidable how close the chosen compression algorithm is to $ \mathbf{K}\left( x \right) $.
This follows from the semi-computability of $ \mathbf{K}\left( \cdot \right) $  \cite{Calude2002,Li1997}:
the exact value of $ \mathbf{K}\left( x \right) $ is not computable, but there are always algorithms that are able to produce better approximations in the asymptotic limit.
When developing new methods of approximating the value of $ \mathbf{K}\left( x \right) $, it is a proved mathematical property that there is always room for improvement with respect to older methods.

Thus, as pointed by \cite{Burgin2009}, algorithmic complexity defines a measure of the minimum information necessary for constructing the object $ x $ with Turing machines, computable processes, or computable functions. 
Indeed, the irreducible information content carried by $ x $ (denoted by the string $ x^* $) with respect to a universal programming language from which $ \mathbf{K}\left( x \right) $ measures its size can only be in general extracted from the encoded object $ x $ by an uncomputable function (in particular, in Turing degree $ \mathbf{ 0' } $).
This also analogously applies to resource-bounded versions of the algorithmic complexity:
the irreducible information content carried by $ x $ in general can only be accessed from $ x $ by processes at a higher computational class than the processes that construct $ x $.
Nevertheless, either in the unbounded or bounded case, the semi-computability of algorithmic complexity assures that the irreducible information content can always be approximated by computational procedures at the same computational class of the processes that construct $ x $.

Therefore, we refer to \emph{algorithmic information content} (a.i.c.) of $ x $ as the minimum necessary and sufficient information for computably constructing $ x $ such that this information can always be extracted from $ x $ by a fixed function at some Turing degree.
We measure the \emph{size $ \mathbf{ I_{ac} }\left( x \right) $ of the algorithmic information content} of $ x $ as the \emph{equivalence class} of integer values $ k \in  \mathbf{ I_{ac} }\left( x \right) $ in the interval 
\begin{equation}
	\left| \mathbf{K}\left( x \right) - k \right| \leq c_\mathbf{I}
	\text{ ,}
\end{equation}
where $ c_\mathbf{I} \in \mathbb{N} $ is an arbitrary and sufficiently large \emph{object-independent} constant.
Besides being fixed and independent of the objects $ x $, the value of the constant $ c_\mathbf{I} $ is sufficiently large to encompass
the other object-independent constants that appear:
in the algorithmic coding theorem; 
in $ \mathbf{ I_A } \left( x ; x^* \right) = \mathbf{ I_A } \left( x^* ; x \right) \pm \mathbf{O}( 1 ) = \mathbf{K}\left( x \right) \pm \mathbf{O}( 1 ) $; 
in $ \mathbf{ I }_{ \mathbf{ K } } \left( x : x \right) = \mathbf{K}\left( x \right) - \mathbf{O}( 1 ) $; 
and in any difference $ \left| \mathbf{ I_{ ac } }\left( x \right) - f\left( x \right) \right| \leq \mathbf{O}(1) $, where $ f\left( \cdot \right) $ is any other arbitrary measure of size of information content that an observer might choose to employ such that $ f\left( \cdot \right) $ is equivalent to $ \mathbf{K}\left( \cdot \right) $.

Also note that the above definition of algorithmic information content
applies analogously to the size $ \mathbf{ I_{ac} }\left( z \, \vert w \right) $ of the conditional algorithmic information content of $ z $ given $ w $, which is an equivalence class of values $ k \in  \mathbf{ I_{ac} }\left( z \, \vert w \right) $ in the interval
\begin{equation}
	\left| \mathbf{K}\left( z \, \vert w \right) - k \right|
	\leq  c_\mathbf{I}
	\text{ ,}
\end{equation}
where the \emph{conditional} prefix algorithmic complexity of a binary string $ z $ given a binary string $ w $, denoted by $ \mathbf{K}( z \, \vert w ) $, is the length of the shortest program $ z_w^*$ such that $ \mathbf{U}( \left< w , z_w^* \right> ) = z $.

The known properties of the algorithmic coding theorem and mutual algorithmic information in AIT are important to note.
This is because the fact that the algorithmic information content measure is a constant-bounded equivalence class $ \mathbf{ I_{ac} }\left( \cdot \right)  $ (instead of a fixed value given by the the algorithmic complexity $ \mathbf{K}\left( \cdot \right) $) directly implies that the values of $ \mathbf{ I_{ac} }\left( \cdot \right) $ in Definitions~\ref{defObserverdependentemergence} and \ref{defAOIE} are \emph{invariant} and \emph{minimal} with respect to a particular observer. 
$ \mathbf{ I_{ac} }\left( x \right) $ is an equivalence class that does not vary by the chosen finite number of: distinct formal methods of measuring irreducible information content; universal programming languages; or encoding methods.
Additionally, as a consequence of the algorithmic coding theorem, the algorithmic probability equivalence class given by $  2^{ - \mathbf{ I_{ac} }\left( x \right)  }  $ is an equivalence class that does not vary by the chosen formal methods of assigning a probability distribution to the infinite discrete space of computably constructible objects.
Thus, as the reader will notice in Section~\ref{sectionEmergence}, although perceiving the behaviour of a particular system as emergent depends on the observer's formal knowledge, one of the important contributions of this article is that the values of algorithmic information content remain invariant and minimal with respect to this observer.

%
%

\subsection{Dynamical systems}\label{sectionAICofCDSs}

For systems composed or defined by stochastic processes, the emergence of information has been studied in terms of statistical information (for example, based on entropy) or other related measures \cite{Prokopenko2009,Fernandez2014}.
Also in the context of stochastic processes, the quantification of synergy in multivariate stochastic systems has been an active field \cite{Lizier2018}. 
However, it is already known that statistics faces insuperable obstacles when attempting to quantify the algorithmic information content of deterministic processes \cite{Zenil2020b}.
This is seen, for example, in
the existence of Borel-normal sequences that are in fact computable (and therefore logarithmically compressible) \cite{Becher2002},
and in the existence of low-algorithmic-complexity graphs in which the degree-sequence entropy is maximal, and also Borel-normal \cite{Zenil2017}.
Thus, if one is interested in measuring algorithmic information content (or measuring the emergence of new irreducible information) in deterministic systems, which are free of stochasticity, employing any computable method only capable of characterising statistical regularities (in order to approximate a compressed form) generally entails these obstacles.
For this reason, the present article only addresses discrete deterministic dynamical systems or computable systems in general.

When moving from the context of encoded objects to discrete dynamical systems, the properties of algorithmic information set forth in Section~\ref{sectionAICofOs} are important to understand in order to grasp the formal sense in which
algorithmic information is a measure of irreducible information content in dynamical systems.
Let $ \mathcal{S} = \left( X_\mathcal{S}  , f_\mathcal{S} , E_\mathcal{S} , T \right) $ be a \emph{finite discrete deterministic dynamical system} (FDDDS) embedded in an environment $ \mathcal{E} $ \cite{Hernandez-Orozco2018}, where $ X_\mathcal{S} $ is the state space of $ \mathcal{S} $, \begin{equation*}
	\begin{array}{lcccc}
		f_\mathcal{S} & \colon & X_\mathcal{S} \times E_\mathcal{S} \times T & \to & X_\mathcal{S} \\
		\, & \, & \left( \mathcal{S}_t , e_{\mathcal{S}_t} , t  \right) & \mapsto & \mathcal{S}_{ t + 1 }
	\end{array}
\end{equation*} 
is the function that defines the \emph{evolution rule} (e.r.) of $ \mathcal{S} $, $  E_\mathcal{S} $ is the space of all possible surrounding environmental states that belong to the boundary of $ \mathcal{S} $, and $ T $ is the set of time instants.
If the cardinality of the set $  E_\mathcal{S} $ of a dynamical system $ \mathcal{S} $ is finite, then the dynamical system is said to have a finite \emph{boundary}.
If both sets $ X_\mathcal{S} $ and $  E_\mathcal{S} $ are composed only of discrete finite states, the finite-boundary dynamical system is said to be \emph{finite} and \emph{discrete}.
In this article we only deal with dynamical systems $ \mathcal{S} $ that are finite, discrete, deterministic, and have finite boundaries.
The environment $ \mathcal{E} = \left( X_\mathcal{E}  , r_\mathcal{E} , T \right) $ is a FDDDS into which the systems $ \mathcal{S} $ and their environmental surroundings $ E_\mathcal{S} $ are embedded, where $ r_\mathcal{E} \colon X_\mathcal{E} \times T \to X_\mathcal{E} $
is the e.r. of $ \mathcal{E} $.
In cases where the e.r. of a dynamical system is a computable function or computable relation, the dynamical system is said to be \emph{computable}.

We define the measure of the \emph{size of the a.i.c. of a FDDDS} $ \mathcal{S} = \left( X_\mathcal{S}  , f_\mathcal{S} , E_\mathcal{S} , T \right) $ until time instant $ t $ by $ \mathbf{ I_{ac} }\left( \mathcal{S} \upharpoonright_0^{ t } \right) $, 
where
$ \mathcal{S} \upharpoonright_0^{ t } $ is simply a notation for an arbitrary encoding of the sequence $ \left(  \mathcal{S}_0 , \mathcal{S}_1 , \dots , \mathcal{S}_t  \right) $ of states (i.e., a state space trajectory of $ \mathcal{S} $ until $ t \in T $).
It is straightforward to show that, with respect to the encoded object $ \mathcal{S} \upharpoonright_0^{ t } $, the equivalence class $ \mathbf{ I_{ac} }\left( \mathcal{S} \upharpoonright_0^{ t } \right) $ inherits the same properties discussed in Section~\ref{sectionAICofOs}.

It is known that TMs can be simulated by computable FDDDSs.
For example, one can construct an \emph{elementary cellular automaton} (ECA) employing Rule 110 that simulates a TM \cite{Wolfram2002}. 
In addition, the decision problem of one is reducible to a decision problem of the other and the time complexity of the TM simulation by ECAs can be improved to a polynomial time overhead \cite{Neary2006}.

\begin{lemma}\label{lemmaFDDDSequivalencewithObjects}
	Let $ \mathcal{S} $ be a \emph{computable} FDDDS such that $ \mathbf{U}\left( \left< \mathcal{S} \upharpoonright_0^{ t } , p \right> \right) = y $ and, for every $ t' < t $, there is $ z $ such that $ \mathbf{U}\left( \left< \mathcal{S} \upharpoonright_0^{ t' } , p \right> \right) = z $ and $ z \neq y $ .
	Then,
	\begin{equation*}
		\left| \mathbf{K}\left( y \right) - \mathbf{K}\left( \mathcal{S} \upharpoonright_0^{ t } \right) \right| \leq \mathbf{K}\left( p \right) + \mathbf{O}\left( 1 \right)
		\text{ .}
	\end{equation*}
	
	\begin{proof}
		From the minimality of $ \mathbf{K}\left( \cdot \right) $, we have it that $ \mathbf{K}\left( y \right) \leq \mathbf{K}\left( \mathcal{S} \upharpoonright_0^{ t } \right) + \mathbf{K}\left( p \right) + \mathbf{O}\left( 1 \right) $.
		Since $ \mathcal{S} $ is computable and $ \mathbf{U}\left( \left< \mathcal{S} \upharpoonright_0^{ t' } , p \right> \right) = z \neq y $ for every $ t' < t $, then we also have it that $ \mathbf{K}\left( \mathcal{S} \upharpoonright_0^{ t } \right) \leq \mathbf{K}\left( y \right) + \mathbf{K}\left( p \right) + \mathbf{O}\left( 1 \right) $.
	\end{proof}
\end{lemma}

Moreover, in case the system $ \mathcal{S} $ is simulating
an arbitrary Turing machine $ w $ and the decision problem of $ \mathcal{S} $ until $ t $ is \emph{Turing equivalent} to the decision problem of $ \mathbf{U}\left( w \right) $, Lemma~\ref{lemmaFDDDSequivalencewithObjects} implies that one can \emph{equivalently measure the a.i.c.} of $ \mathcal{S} $ by $ \mathbf{ I_{ac} }\left( y \right) $ instead of $ \mathbf{ I_{ac} }\left( \mathcal{S} \upharpoonright_0^{ t } \right) $, where $ \mathbf{U}\left( w \right) = y $. And the conditional case $ \mathbf{ I_{ac} }\left( \cdot \middle\vert \cdot \right) $ applies analogously.

We can now obtain from Lemma~\ref{lemmaUncomputabilityofS} another property related to the predictability or computability of FDDDSs that will be important to the results in Section~\ref{sectionEmergence}.
As usual, let $f(x)=\mathbf{o}( g(x) )$ denote function $g$ \emph{strongly} dominating function $f$ asymptotically.

\begin{lemma}\label{lemmaUncomputabilityofS}
	Let $ \mathcal{S} $ be a FDDDS for which 
	\begin{equation*}
		\mathbf{K}\left( A \right) + \mathbf{K}\left( B \right) + \log\left( t \right) = \mathbf{o}\left( \mathbf{ K }\left( \mathcal{S} \upharpoonright_0^{ t } \right) \right)
		\text{ ,}
	\end{equation*}
	where $ \mathbf{K}\left( A \right) $ and $ \mathbf{K}\left( B \right) $ are the algorithmic complexity of encoded finite subsets $ A \subset X_\mathcal{S} $ and $ B \subset E_\mathcal{S} $, respectively.
	Then, for every program $ p $ and for every formal axiomatic theory $ \mathbf{F} $, there is $ t' $ such that, for every $ t \geq t' $,
	\begin{equation*}
		\mathbf{U}\left( \left< t, \mathbf{F} , A , B , p \right> \right) \neq \mathcal{S} \upharpoonright_0^{ t }
		\text{ .} 
	\end{equation*}
	
	\begin{proof}
		The proof directly follows by contradiction of the fact that, for any fixed $ p $ and $ \mathbf{F} $, if 
			$ \mathbf{U}\left( \left< \left| s \right| , \mathbf{F} , A , B , p \right> \right) = s
			\text{ ,} $
		then $ \mathbf{ K }\left( s \right) \leq \mathbf{O}\left( \log\left( \left| s \right| \right) \right) + \mathbf{K}\left( A \right) + \mathbf{K}\left( B \right) + \mathbf{O}( 1 ) $.
	\end{proof}
\end{lemma}

In this regard, \citet{Hernandez-Orozco2018} achieves a more general result.
The notion of a system adapting to the environment at infinitely many time instants is formalised in \cite{Hernandez-Orozco2018} as \emph{weak convergence} of $ \mathcal{S} $ toward $ \mathcal{E} $, i.e., for an $ \epsilon > 0 $, there is an infinite set $ T' \subset T $ such that, for every $ t \in T' $,
	$ \mathbf{K}\left( \mathcal{E}_{ t + 1 }  \vert \mathcal{S}_{ t + 1 }  \right) \leq \epsilon
	\text{ .} $
If $ \mathcal{S} $ displays \emph{strong Open-Endedness}, 
\citet{Hernandez-Orozco2018} demonstrates that one cannot even decide at which time instants the system $ \mathcal{S} $ is adapting to the environment.

\section{Observers and algorithmic perturbations}\label{sectionObserversystems}


\subsection{Algorithmic perturbations}\label{sectionAlgorithmicperturbations}

Intuitively, an observation of $ \mathcal{S} $ by the observer should be realised when the interaction between them somehow sends sufficient information about $ \mathcal{S} $ to the observer.
To formally tackle this problem within the scope of AID, any interaction between systems is reduced to a number of algorithmic perturbations.

Since any state space trajectory of a FDDDS is a sequence of finite discrete states, then, for every perturbation of a state $ \mathcal{S}_{ t } $ at time instant $ t $ that results into the next state $ \mathcal{S}'_{ t + 1 } $, there is at least one computer program (or Turing machine) that performs this exact change by taking $ \mathcal{S}_{ t } $ as input and outputting $ \mathcal{S}'_{ t + 1 } $.
If a state space trajectory of a dynamical system produces a sequence in which the state $ A_{ t + 1 } $ should occur one time step after the state $ A_{ t } $ and, instead, another state $ A'_{ t + 1 } $ is occurring one time step after $ A_{ t } $, then a perturbation has changed the original one-time-step course of the state space trajectory of the dynamical system.
In addition, if the states of the dynamical system are finite and discrete for any time instant $ t $, then there is at least one program, Turing machine, or computable process that can output a state $ A'_{ t + 1 } $ from the past state $ A_{ t } $ as input.
Those programs, Turing machines, or computable processes that change the original one-time-step course of the state space trajectory of a finite discrete dynamical system are called algorithmic perturbations.

This way, an \emph{algorithmic perturbation} is defined as such a program that changes the course of the state space trajectory of a FDDDS by taking the previous (not yet perturbed) state of this FDDDS as input and outputting  another future state, which is distinct from what the next state of the FDDDS should be if \emph{no} perturbation had had happened.
Intuitively, an algorithmic perturbation occurring at a certain time instant changes the course of the state space trajectory from that moment on.
In other words, an algorithmic perturbation occurring at time instant $t$ is any kind of external algorithmic process that updates the state of the affected system after one time step, resulting into a new state at time instant $t+1$.
That is, instead of the state $ \mathcal{S}_{ t + 1 } $ that $ \mathcal{S} $ should display if no perturbation had had occurred, $ \mathcal{S} $ displays $ \mathcal{S}'_{ t + 1 } $ after an algorithmic perturbation has occurred at time instant $t$.
For example, suppose a \emph{non}-perturbed FDDDS $ \mathcal{S} $ at time instant $ t $ displays state $ \mathcal{S}_t $ and, at time instant $ t + 1 $, it displays state $ \mathcal{S}_{ t + 1 } $.
Now, if an algorithmic perturbation affects $ \mathcal{S} $ at time instant $t$, then it takes $ \mathcal{S}_t $ as input and returns a state $ \mathcal{S}'_{ t + 1 } $ to be displayed by $ \mathcal{S} $ at time instant $t+1$, state which is distinct from the state $ \mathcal{S}_{ t + 1 } $ that should be displayed at time instant $t+1$ if the algorithmic perturbation had \emph{not} had occurred in first place.
As more concrete example, suppose each state of a dynamical system $ A $ is a $3$-bit string and suppose $ A_t = 001 $ and $ A_{ t + 1 } = 010 $.
If a perturbation occurs at time instant $ t $ and it leads the next state to be $ A'_{ t + 1 } = 011 $ instead of $ A_{ t + 1 } = 010 $, we know there is at least one algorithm that corresponds to this exact perturbation.
In particular, such an algorithm can be as simple as ``read the first two bits of the input and flip the second bit, then returns the resulting 3-bit string'', where
$ \mathcal{P}_2 $ is a program of a Turing machine that represents this algorithm so that $ \mathbf{U}\left( \left< 001 , \mathcal{P}_2 \right> \right) = 011 $.

\begin{definition}\label{defAlgorithmicperturbation}
	An \emph{algorithmic perturbation} (AP) $ \mathcal{P} $ at time instant $ t $ is a perturbation occurring at the time instant $ t $ of a state space trajectory $ \left( \dots , A_t \right) $ of a finite discrete dynamical system $ A $ that updates the one time step from $ t $ to $ t + 1 $ so that, instead of the original state space trajectory $ \left( \dots , A_t , A_{ t + 1 } , \dots \right) $, it results in a distinct state space trajectory $ \left( \dots , A_t , A'_{ t + 1 } , \dots \right) $, where $ \mathcal{P} $ is a program and $ \mathbf{U}\left( \left< A_t , \mathcal{P} \right> \right) = A'_{ t + 1 } $.
\end{definition}


Note that the following Definition~\ref{defAlgorithmicperturbation} applies, with no loss of generalisation, to finite discrete dynamical systems.
FDDDSs are particular cases of finite discrete dynamical systems and computable FDDDSs are particular cases of FDDDSs.

Note that the existence of an algorithmic perturbation does not depends on where it comes from or which is the nature of the process that caused the perturbation.
What mathematically follows from the definition of algorithmic perturbation is simply that: any finite state change in a finite discrete dynamical system $ A $ can be reduced to, or represented by, an equivalent algorithmic perturbation into $ A $; and that any halting program $ \mathcal{P} $ on input $ A_t $ is a possible algorithmic perturbation that may (or may not) occur on system $ A $ at time instant $ t $.
Whether or not one is assigning probabilities to the occurrence of perturbations depends on the problem and model to be studied.
From the example in the previous paragraph, suppose one knows beforehand that the second bit was flipped due to a stochastically random event (with probability $ 1/3 $) in which the bit to be flipped is selected randomly.
But, since the result of such a perturbation just produces $ 011 $ out of $ 001 $, then that change can be equivalently represented by the algorithmic perturbation $ \mathcal{P}_2 $ or any other algorithmic perturbation $ \mathcal{P} $ such that $ \mathbf{U}\left( \left< 001 , \mathcal{P} \right> \right) = 011 $.
Because any program can play the role of an algorithmic perturbation, there are always an infinite denumerable number of $ \mathcal{P} $ such that $ \mathbf{U}\left( \left< 001 , \mathcal{P} \right> \right) = 011 $.
Thus, if $ \mathcal{P}_2 $ is the algorithmic perturbation one chooses (possibly, because $ \left| \mathcal{P}_2 \right| $ is minimal) to represent the stochastically random event of flipping the second bit, which is a stochastic perturbation that occurs with probability $ 1/3 $, then the probability of occurrence of the algorithmic perturbation $ \mathcal{P}_2 $ also becomes $ 1/3 $. 

Note that we are employing the term `stochastic randomness' to distinguish it from algorithmic randomness and avoid ambiguities. 
A stochastically random event is one produced by stochastic processes.
Thus, a stochastically random perturbation is a perturbation that changes (i.e., deletes or inserts) the elements of a system according to a probability distribution \cite{Zenil2020,Zenil2020a,Zenil2019a}.
On the other hand, an algorithmically random encoded object (or event) is the one that is incompressible \cite{Li1997,Downey2010,Calude2002,Chaitin1987}.
Additionally, note that not every algorithmically random object is generated by a stochastic process.
For example, the halting probability \cite{Chaitin1987} is an incompressible number, while there is a deterministic (but uncomputable) process that generates it (in particular, a function in Turing degree $ \mathbf{ 0' } $).

While the conditional algorithmic complexity of any perturbation that results in $ 011 $ from the past state $ 001 $ is constant and very small---because the simplest algorithmic perturbation $ \mathcal{P} $ for which $ \mathbf{U}\left( \left< 001 , \mathcal{P} \right> \right) = 011 $ holds can only be as complex as flipping the second bit---, finding the algorithmic complexity of the equivalent algorithmic perturbations to stochastically random perturbations on more complex objects, such as networks, is less trivial.
In the case of monoplex networks (or graphs), it is shown in \cite{Zenil2020,Zenil2020c} that stochastic-randomly deleting (or inserting) $ \left| F \right| $ edges in a network $ G $, which results in a new network $ G' $, is equivalent to applying an algorithmic perturbation $ \mathcal{P}_F $ on $ G $ such that 
\begin{equation}\label{equationMILSresult1}
\mathbf{K}\left( \mathcal{P}_F \right) \leq 2 \left| F \right| \log_2\left( N \right) +  \mathbf{O}\left(\log_2 \left( \left| F \right| \right)\right) + \mathbf{O}( \log_2 \left(\log_2 ( N )\right) )
\end{equation}
and $ \mathbf{U}\left( \left< G , \mathcal{P}_F \right> \right) = G' $, where $ F $ is the subset of edges that were perturbed and $ N $ is the number of vertices.
This is because the right side of the inequality in Equation~\ref{equationMILSresult1} is an upper bound for the conditional algorithmic complexity of the shortest program that, with the network $ G $ as input, can perform the same edge deletions (or edge insertions) that the stochastically random deletion (or insertion) of $ \left| F \right| $ edges did.

For example, suppose the probability that a destructive stochastic perturbation deletes a \emph{single} edge is $ \frac{ 2 }{ N^2 - N } $.
We know there is an equivalent algorithmic perturbation $ \mathcal{P}_{F_1} $ such that $ \mathbf{K}\left( \mathcal{P}_{F_1} \right) \leq 2 \log_2\left( N \right) + \mathbf{O}( \log_2 \left(\log_2 ( N )\right) ) $ \cite{Zenil2020,Zenil2020c}. 
If one chooses $ \mathcal{P}_{F_1} $ to represent that exact stochastically random deletion, then the probability of occurrence of $ \mathcal{P}_{F_1} $ will also be $ \frac{ 2 }{ N^2 - N } $. 
Indeed, as one of the important properties implied by this equivalence in algorithmic information dynamics, a stochastically random perturbation on a single edge can only change the final algorithmic complexity of the network by $ \mathbf{O}( \log_2 (N) ) $ bits, which explains the thermodynamic-like behaviour found in \cite{Zenil2019a} about the reprogrammability of networks when these are subjected to stochastically random single-edge perturbations. 
More specifically, this thermodynamic-like phenomenon refers to a larger number of one-by-one stochastically random edge deletions (or insertions) being necessary for transforming an algorithmically random (i.e., incompressible) network into a low-algorithmic-complexity network than the number necessary for transforming a low-algorithmic-complexity network into an algorithmically random network.
This is because (stochastically) random single-edge deletion on algorithmic simple networks has a greater impact than (stochastically) random single-edge deletion on algorithmic random networks \cite{Zenil2019a}.  
Otherwise said, to lower the algorithmic randomness of an algorithmically random graph, non-stochastic single-edge deletion is required.
However, to make a low algorithmic complexity network a higher algorithmic complexity network, stochastically random single-edge deletion suffices \cite{Zenil2020c}. 
Moreover, single-edge (or single-node) deletion on an algorithmically random or simple network (e.g., a complete graph) has a much smaller impact on the resulting algorithmic complexity of the network than a single-edge (or single-node) deletion on neither low nor high algorithmic complex network, therefore, constituting a measure of sophistication that evinces the existence of a reprogrammability optimum point between maximal and minimum algorithmic complexity of a network \cite{Zenil2019a}.

\subsection{Formal observer systems}\label{sectionFOS}

The act of observing is an act of a system (the observer) perturbing an object (another system), while being itself perturbed by the object.
Thus, following the framework in Section~\ref{sectionAICofCDSs}, we define a \emph{formal observer system} (FOS) as a particular type of FDDDS $ \mathcal{S} $, which we will denote by $ \mathcal{O} $ in order to avoid confusion with the general FDDDSs denoted by $ \mathcal{S} $. 
The mathematical challenge is then to establish the necessary conditions for this mutual perturbation to result in sufficient knowledge acquisition by the observer.  
Hence we first introduce a variation of Turing machines (TMs), called \emph{observer Turing machines}, that are going to be simulated by the FOSs.
Then we present the \emph{observation principles}.

An \emph{observer Turing machine} (OTM) $ O $ is a slight variation of the usual 2-tape Turing machine \cite{Lewis1997,Rich2007}, where $ O( x_1 , x_2 ) $ denotes the output of $ O $ with $ x_1 $ in the first tape and $ x_2 $ in the second tape.
The \emph{first tape} works as the first tape of the 2-tape Turing machine, but the \emph{second tape} is filled with an encoded form of a \emph{formal axiomatic theory} (FAT) $ \mathbf{F} $ before the OTM starts to run.
In other words, the OTM is a 2-tape Turing machine with access to a finite-length oracle, and this oracle is precisely the encoding of a FAT, where  $ O( x , \mathbf{F} ) $ denotes the output of $ O $ with input $ x $ when the formal theory $ \mathbf{F} $ is previously known.

Since $ \mathbf{F} $ is always finite, it is straightforward to show that, for every OTM, there is an equivalent single-tape Turing machine that simulates the OTM.
Let $ M $ be a \emph{non-halting} TM that repeatedly simulates an OTM such that: 
$ M $ does not halt if the OTM does not halt; and $ M $ indefinitely repeats the simulation of the OTM from the beginning if the OTM halts. 
Moreover, we know that one can construct a FDDDS that simulates $ M $ (see Section~\ref{sectionAICofCDSs}).
This universal FDDDS can be constructed so that the decision problem of the OTM is Turing equivalent to the decision problem of the universal FDDDS.
As a consequence, one can define a formal observer system as a FDDDS that simulates $ M $ in which the set $ T $ is infinite:

\begin{definition}\label{defFOS}
	A \emph{formal observer system} (FOS)  $ \mathcal{O} $ is defined as a FDDDS that simulates the non-halting TM $ M $ which repeatedly simulates an OTM. 
\end{definition}

Thus, $ \mathcal{O} $ is a FDDDS that extends indefinitely in time either because the OTM does not halt or because it is simulating a halting OTM in an infinite number of repeated \emph{simulation cycles}.
In particular, if the OTM halts, then $ \mathcal{O} $ is a FDDDS whose \emph{recurrence time} \cite{Adams2017} (see also Section~\ref{sectionExamplesofODE}) corresponds to the time steps necessary for the simulation of the OTM by the FDDDS.
In other words, if the OTM halts, simulation cycles correspond to contiguous recurrent state space trajectories.

The construction of the FOSs allows us to formalise in Definition~\ref{defComputabilitybyO} the notion of a function being computed by a state space trajectory, which will be proved to be satisfiable in Lemma~\ref{lemmaSequence computabilityofO}.

\begin{definition}\label{defComputabilitybyO}
	Let $ \mathcal{O} $ be a FOS and let $ f $ be a computable function.
	We say $ \left( \mathcal{O}_{ 0 } , \dots , \mathcal{O}_{ t } , \dots , \mathcal{O}_{ t' } , \dots \right) $ \emph{computes} the value of $ f\left( x \right) $ (or \emph{decides} a problem) between time instants $ t $ and $ t' $ iff the state space trajectory $ \left( \mathcal{O}_{ t } , \dots , \mathcal{O}_{ t' }  \right) $ corresponds to the simulation cycle of the OTM that computes the function $ f $ with input $ x $ (or decides the problem).
\end{definition}

This way, except for the simulation running time (or space) overhead, the (time or space) \emph{computational class} of $ \mathcal{O} $ becomes the (time or space) complexity class of the OTM. 

A direct consequence of Definitions~\ref{defAlgorithmicperturbation},~\ref{defFOS}, and \ref{defComputabilitybyO} is that they enable an AP to change the machine that $  \mathcal{O}  $ is simulating.
If an AP at time instant $ t $ replaces $ \mathcal{O}_{ t + 1 } $ with another state $ \mathcal{O}'_{ t + 1 } $ such that $ \mathcal{O}'_{ t + 1 } $ is the initial state of the universal FDDDS that simulates the same OTM of $  \mathcal{O}  $ but with the first tape containing $ w $, 
then the state space trajectory
$ \left(  \mathcal{O}'_{ t + 1 }, \mathcal{O}'_{ t + 2 } , \dots \right) $ simulates the OTM with input $ w $ in the first tape.
If the OTM with input $ w $ computes a function $ f\left( w \right) $, then $ \left(  \mathcal{O}'_{ t + 1 }, \mathcal{O}'_{ t + 2 } , \dots \right) $ also computes a function $ f\left( w \right) $ after $ t $ time steps.
In the interests of concision, we refer to the first tape and the second tape of the OTM that $  \mathcal{O}  $ is simulating as the \emph{first tape} and \emph{second tape} of $  \mathcal{O}  $, respectively.

For example, suppose $ \left( \mathcal{O}_{ 0 } , \dots , \mathcal{O}_{ t } \right) $ is a state space trajectory that computes the value of $ f\left( w_1 \right) $ with $ w_1 $ as input in its first tape, where $ f $ is a total computable function.
Note that the observer Turing machine that the FDDDS $ \mathcal{O} $ is simulating computes the value of $ f\left( w_1 \right) $ when $ w_1 $ is encoded in its first tape.
So, from our supposition, $ \left( \mathcal{O}_{ 0 } , \dots , \mathcal{O}_{ t } \right) $ corresponds to the state space trajectory that simulates this observer Turing machine with $ w_1 $ in its first tape.
Then, suppose that an algorithmic perturbation occurs at time instant $ t $, giving rise to the next state $ \mathcal{O}'_{ t + 1 } $ instead of the original state $ \mathcal{O}_{ t + 1 } $, which was supposed to occur if no algorithmic perturbation had had happened at time instant $ t $,
so that $ \mathcal{O}'_{ t + 1 } $ is the initial state of the FDDDS that simulates the same observer Turing machine of $  \mathcal{O}  $ but with the first tape containing $ w_2 $. 
So, note that the FDDDS $ \mathcal{O} $ after the algorithmic perturbation has occurred at time instant $ t$ will be simulating the observer Turing machine that computes the value of $ f\left( w_2 \right) $ when $ w_2 $ is encoded in its first tape.
As a consequence, the state space trajectory
$ \left(  \mathcal{O}'_{ t + 1 }, \mathcal{O}'_{ t + 2 } , \dots \right) $ will be simulating the observer Turing machine with input $ w_2 $ in the first tape.
Therefore, $ \left( \mathcal{O}_{ 0 } , \dots , \mathcal{O}_{ t } \right) $ computes the value of $ f\left( w_1 \right) $ and $ \left(  \mathcal{O}'_{ t + 1 }, \mathcal{O}'_{ t + 2 } , \dots \right) $ computes the value of $ f\left( w_2 \right) $.

All of the above results, remarks, and examples lead us to the following Lemma~\ref{lemmaSequence computabilityofO} about FOSs that can receive external information (i.e., by being algorithmically perturbed) over time. 
Lemma~\ref{lemmaSequence computabilityofO} also demonstrates that Definition~\ref{defComputabilitybyO} is satisfiable.

\begin{lemma}\label{lemmaSequence computabilityofO}
	For every $ w_1, w_2 , p $ with $ \mathbf{U}\left( \left< w_1 , p \right> \right) = y_1 $ and $ \mathbf{U}\left( \left< w_2 , p \right> \right) = y_2 $, there is a FOS $ \mathcal{O} $, an AP $ \mathcal{P} $, and a time instant $ t \in T $ such that
	$ \left( \mathcal{O}_{ 0 } , \dots , \mathcal{O}_{ t } , \mathcal{O}'_{ t + 1 }, \mathcal{O}'_{ t + 2 } , \dots \right) $ computes $ y_1 $ until time instant $ t $ and computes $ y_2 $ after time instant $ t $.
	
	\begin{proof}
		Let $ A $ be a universal FDDDS that simulates any TM $ p $.
		Let $ M' $ be a single-tape TM that simulates a 2-tape TM with $ x_1 $ in the first tape and $ x_2 $ in the second tape.
		Let $ M $ be the TM that receives any TM $ p' $, simulates $ p' $ and: 
		if $ p' $ reaches a halting state, writes $ p' $ on the tape again with the head on the first cell at the same initial state as $ M $;
		otherwise, the simulation of $ p' $ continues indefinitely.
		Therefore, we have it that $ M\left( p' \right) $ is undefined for every $ p' $.
		Now, let $ \mathcal{O} $ be a FDDDS that has the same e.r. as $ A $ and an initial state $ \mathcal{O}_0 $ that corresponds to the initial configuration of $ A $ simulating
		$ M\left( M'\left( w_1 , p \right) \right) $.
		Then,  $ \mathbf{U}\left( \left< w_1 , p \right> \right) = y_1 $ implies that there are $ t' \geq 0 $ and $ k \geq 1 $ such that $ \left( \mathcal{O}_{ 0 } , \dots , \mathcal{O}_{ t' }  \right) $ computes $ y_1 $ and $ \mathcal{O}_{ t' + k } = \mathcal{O}_{ 0 } $.
		Let $ \mathcal{O}'_{ t' + k } $ be the initial state that corresponds to the initial configuration of $ A $ simulating
		$ M\left( M'\left( w_2 , p \right) \right) $.
		Since there are only finite states, we have it  that there is an AP $ \mathcal{P} $ at time instant $ t = t' + k - 1 $ such that $ \mathbf{U}\left( \left< \mathcal{O}_{ t } , \mathcal{P} \right> \right) = \mathcal{O}'_{ t' + k } $.
		Therefore, the resulting state space trajectory $ \left( \mathcal{O}_{ 0 } , \dots , \mathcal{O}_{ t } , \mathcal{O}'_{ t + 1 }, \mathcal{O}'_{ t + 2 } , \dots \right) $ computes $ y_1 $ until time instant $ t $ and computes $ y_2 $ after time instant $ t $.
	\end{proof}
\end{lemma}

As an illustrative example of the fact that certain functions (or languages) can be computed (or decided) by a formal observer system while another formal observer system cannot, suppose that Peano arithmetics is consistent (i.e., it does not prove contradictions) and that a formal observer system simulates a two-tape Turing machine whose first tape is empty and the second tape contains an encoding of the axioms of Peano arithmetics, so that this Turing machine tries to prove whether or not the received input in the first tape is a true arithmetical sentence by only using Peano arithmetics.
Suppose now that an algorithmic perturbation causes the arithmetical sentence ``$ Con(PA) $'' (which asserts the consistency of Peano arithmetics) to be encoded into the first tape of such a formal observer system.
That is, this perturbation replaces the next state of the observer so that, instead of corresponding to a state of a (non-perturbed) FDDDS that simulates the exact former two-tape Turing machine, it corresponds to a state of a (perturbed) FDDDS that simulates the former two-tape Turing machine, but with the first tape filled out with a bit string which encodes the sentence ``$ Con(PA) $''.
Then, we know that two-tape Turing machine will never halt and, therefore, the formal observer system will necessarily  be simulating a \emph{non-halting} program.
This holds because of the incompleteness of Peano arithmetics.
Now, suppose that another formal observer system simulates a two-tape Turing machine whose first tape is empty and has the axioms of Peano arithmetic plus the extra axiom ``$ Con(PA) $'' encoded into its second tape.
Then, in case an algorithmic perturbation causes the arithmetical sentence ``$ Con(PA) $'' to be encoded into the first tape of such a formal observer system, the formal observer system will be simulating a \emph{halting} program that proves the consistency of Peano arithmetic.

By generalising such a mathematical property, whether or not a formal observer system can compute a certain function is a fact dependent on the prior formal knowledge that the formal observer system knows.
This is an important property that we will explore in the observer-dependent emergence.

\subsection{Observation principle}\label{sectionObservationprinciple}

Definition~\ref{defObservationprinciple} formalises the principles necessary for a proper observation to occur between a formal observer system $ \mathcal{O} $ and the FDDDS $ \mathcal{S} $. 
The main idea is that an observation occurs when both the formal observer system $ \mathcal{O} $ and the FDDDS $ \mathcal{S} $ are algorithmically perturbing each other in such a way that the algorithmic perturbation from $ \mathcal{S} $ into $ \mathcal{O} $, which occurs at time instant $ t $, introduces sufficient information about $ \left( \mathcal{S}_{ t - k } , \dots , \mathcal{S}_{ t } \right) $ and $ \mathcal{S}'_{ t + 1 } $ in the first tape of $ \mathcal{O} $, so that this information is contained in (or encoded into) the immediate next state $ \mathcal{O}'_{ t + 1 } $ of the formal observer system.

The condition that the algorithmic perturbation needs to supply information about the post-perturbation future state $ \mathcal{S}'_{ t + 1 } $ (see Equation~\ref{equationObservationprinciple}) is necessary because the act of observing (which is a mutual algorithmic perturbation occurring at time instant $ t $) is only realised by a state of the formal observer system at time instant $ t + 1 $, which is exactly one time step after the algorithmic perturbations have happened.
Thus, at the right moment that a successful observation is realised, the observer must have acquired sufficient information not only about the observed system, but also about what changes the very act of observing introduced into the observed system at the exact moment that the observation happened.
Otherwise, $ \mathcal{O}'_{ t + 1 } $ could have acquired sufficient algorithmic information about $ \left( \mathcal{S}_{ t - k } , \dots , \mathcal{S}_{ t } \right) $ to compute $ \left( \mathcal{S}_{ t - k } , \dots , \mathcal{S}_{ t } , \mathcal{S}_{ t + 1 }  \right) $, while not having acquired sufficient information to compute $ \mathcal{S}'_{ t + 1 } $.
In other words, the algorithmic perturbations in the act of observing could have been informative about the past of the observed system despite not being sufficient to determine what changes the very act of observing caused in the observed system.
Informally, this means that (after the algorithmic perturbation in the act of observing) the system being observed would actually become substantially distinct from what the observer ``thinks'' the system should be, and therefore the act of observing would not have given the observer sufficient information about the system being observed. 

Analogously to $ \mathcal{S} \upharpoonright_0^{ t } $ in Section~\ref{sectionAICofCDSs}, let $ \mathcal{S} \upharpoonright_{ t }^{ t' } $ denote an arbitrarily chosen encoding of the state space trajectory $ \left( \mathcal{S}_t , \dots , \mathcal{S}_{ t' } \right) $ of  $ \mathcal{S} $ between time instants $ t $ and $ t' $.
In short, let $ \left( \mathcal{S} \upharpoonright_{ t - k }^{ t } ,  \mathcal{S}'_{ t + 1 } \right) $ denote the state space trajectory $  \left( \mathcal{S}_{ t - k } , \dots , \mathcal{S}_{ t } , \mathcal{S}'_{ t + 1 } \right) $ and $ \left< \mathcal{S} \upharpoonright_{ t - k }^{ t } ,  \mathcal{S}'_{ t + 1 } \right> $ denote any arbitrarily chosen encoding of the sequence $ \left( \mathcal{S} \upharpoonright_{ t - k }^{ t } ,  \mathcal{S}'_{ t + 1 } \right) $.

Note that, since any formal observer system is simulating an observer Turing machine, then the algorithmic information content in the state space trajectory $ \left( \mathcal{O} \upharpoonright_{ 0 }^{ t } , \mathcal{O}'_{ t + 1 } \right) $ is equivalent to the one contained in the string $ \left< w , \mathcal{O} \upharpoonright_{ 0 }^{ t } \right> $, where $ \mathcal{O}'_{ t + 1 } = \mathbf{U}\left( \left< \mathcal{O}_t , \mathcal{P}_{ \left( \mathcal{S} , \mathcal{O} , t \right) } \right> \right) $ and $ w $ is the string in the first tape of the observer Turing machine that the state space trajectory $ \left(  \mathcal{O}'_{ t + 1 }, \mathcal{O}'_{ t + 2 } , \dots \right) $ is simulating after the occurrence of the algorithmic perturbation $ \mathcal{P}_{ \left( \mathcal{S} , \mathcal{O} , t \right) } $.
Thus, the reader may interchangeably replace Equation~\ref{equationObservationprinciple} with $ \mathbf{K}\left(  \left< \mathcal{S} \upharpoonright_{ t - k }^{ t } ,  \mathcal{S}'_{ t + 1 } \right> 
\middle\vert 
\left< \mathcal{O} \upharpoonright_{ 0 }^{ t } , \mathcal{O}'_{ t + 1 } \right> \right) \leq c_\mathcal{O}  $ without loss of generality.
In Definition~\ref{defObservationprinciple}, we chose to write Equation~\ref{equationObservationprinciple} by employing $ \left< w , \mathcal{O} \upharpoonright_{ 0 }^{ t } \right> $ instead of $ \left< \mathcal{O} \upharpoonright_{ 0 }^{ t } , \mathcal{O}'_{ t + 1 } \right> $ in order to facilitate the intuitive understanding that: some algorithmic information about the past $ k + 1 $ states of $ \mathcal{S} $ and about the immediate post-perturbation state $ \mathcal{S}'_{ t + 1 } $ need to be introduced into the formal observer system by the algorithmic perturbation $ \mathcal{P}_{ \left( \mathcal{S} , \mathcal{O} , t \right) } $; 
and $ w $ can be interpreted as a ``representation'' (of the past $ k + 1 $ states of $ S $ and of the state $ \mathcal{S}'_{ t + 1 } $) that contains such a sufficient amount of algorithmic information.

In short, we say an AP at time instant $ t $ \emph{causes} a bit string $ w $ to be encoded into the first tape of $ O $ (which is simulated by the FOS $ \mathcal{O} $) iff $ w $ is encoded into the first tape of the observer Turing machine that $ \left(  \mathcal{O}'_{ t + 1 }, \mathcal{O}'_{ t + 2 } , \dots \right) $ is simulating and $ \mathcal{O}'_{ t + 1 } $ is the output of this AP.

\begin{definition}[Observation principle]\label{defObservationprinciple}
	\noindent
	Let $ O $ be the OTM that $ \mathcal{O} $ is simulating.
	Let $ c_\mathcal{O} \in \mathbb{N} $ be a constant that only depends on $ \mathcal{O} $ and does not depend on $ \mathcal{S} $.
	Let $ \mathcal{P}_{ \left( \mathcal{O} , \mathcal{S} , t \right) } $ be an AP from $ \mathcal{O} $ into $ \mathcal{S} $ at time instant $ t $.
	Let $ \mathcal{P}_{ \left( \mathcal{S} , \mathcal{O} , t \right) } $ be an AP from $ \mathcal{S} $ into $ \mathcal{O} $ at time instant $ t $.
	We say $  \mathcal{O}  $ \emph{observes} the past $ k + 1 $ states of $ \mathcal{S} $ at time instant $ t $ 
	if $ \mathcal{P}_{ \left( \mathcal{S} , \mathcal{O} , t \right) } $ \emph{causes} a bit string $ w $ to be encoded into the first tape of $ O  $ such that
	\begin{equation}\label{equationObservationprinciple}
	\mathbf{K}\left(  \left< \mathcal{S} \upharpoonright_{ t - k }^{ t } ,  \mathcal{S}'_{ t + 1 } \right> 
	\middle\vert 
	\left< w , \mathcal{O} \upharpoonright_{ 0 }^{ t } \right> \right) \leq c_\mathcal{O} 
	\text{ ,}
	\end{equation}
	\noindent where $ \mathcal{S}'_{ t + 1 } = \mathbf{U}\left( \left< \mathcal{S}_t , \mathcal{P}_{ \left( \mathcal{O} , \mathcal{S} , t \right) } \right> \right) $.

\end{definition}

It follows directly from the basic properties in algorithmic information theory that the constant $ c_\mathcal{O} $ sets the error margin for the extent to which the mutual algorithmic information between the system being observed and the information obtained by the observer during the observation is preserved.
The smaller the value of $ c_\mathcal{O} $, the more \emph{mutual algorithmic information} (and also $ \mathbf{K} $-complexity of information) is preserved.
Formally, if the Definition~\ref{defObservationprinciple} is satisfied, then we have it that
\begin{equation}\label{equationMutualinformationpreserved}
	\begin{aligned}
		\mathbf{K}\left( \left< \mathcal{S} \upharpoonright_{ t - k }^{ t } ,  \mathcal{S}'_{ t + 1 } \right> \right) - c_\mathcal{O}
		& \leq 
		\mathbf{ I_K }\left( 
		\left< w , \mathcal{O} \upharpoonright_{ 0 }^{ t } \right>
		:
		\left< \mathcal{S} \upharpoonright_{ t - k }^{ t } ,  \mathcal{S}'_{ t + 1 } \right> \right)
		\leq \\
		& \leq 
		\mathbf{ I_A }\left( 
		\left< w , \mathcal{O} \upharpoonright_{ 0 }^{ t } \right>
		;
		\left< \mathcal{S} \upharpoonright_{ t - k }^{ t } ,  \mathcal{S}'_{ t + 1 } \right> \right) + \mathbf{O}( 1 )
		\leq \\
		& \leq
		\mathbf{K}\left( \left< \mathcal{S} \upharpoonright_{ t - k }^{ t } ,  \mathcal{S}'_{ t + 1 } \right> \right) + \mathbf{O}( 1 )
		\text{ .}
	\end{aligned}
\end{equation}
Thus, the act of observing formalised in Definition~\ref{defObservationprinciple} is general enough to encompass the case in which observation takes place, but it is defective.
That is, when $ \mathcal{O} $ observes $ \mathcal{S} $ at time instant $ t $ and it only obtains partial information about $ \mathcal{S} $.
In other words, this defective information about $ \mathcal{S} $ can differ from the actual information about $ \mathcal{S} $, but only up to a bounded error margin (given by the constant $ c_\mathcal{O} $ that does not depend on $ \mathcal{S} $ and only depends on $ \mathcal{O} $),
which instantiates and delimits the subjective nature of the act of observing.
Intuitively, this observation may be only acquiring partial information, and not all the desired information, due to: either intrinsic limitations of the properties of the formal observer system, such as limited sensory capabilities or measurement accuracy; a stronger effect of the algorithmic perturbation $ \mathcal{P}_{ \left( \mathcal{O} , \mathcal{S} , t \right) } $; or both.
In all of such examples, it is evinced the subjective character of the formal observer system, subjectivity which is reflected on the value of the constant $ c_\mathcal{O} $.

In addition to defective observations, in some cases the observation can be \emph{ideal} or \emph{perfect}.
Let $ O $, $ \mathcal{O} $, $ \mathcal{P}_{ \left( \mathcal{O} , \mathcal{S} , t \right) } $, and $ \mathcal{P}_{ \left( \mathcal{S} , \mathcal{O} , t \right) } $ be as in Definition~\ref{defObservationprinciple}.
Let $ p $ be program that does not depend on $ \mathcal{O} $ nor $ \mathcal{S} $.
We say $  \mathcal{O}  $ \emph{perfectly observes} the past $ k + 1 $ states of $ \mathcal{S} $ at time instant $ t $ 
if $ \mathcal{P}_{ \left( \mathcal{S} , \mathcal{O} , t \right) } $ causes a bit string $ w $ to be encoded into the first tape of $ O  $ such that $ \mathbf{U}\left( \left< w , \mathbf{F} , O , p \right> \right) = \left< \mathcal{S} \upharpoonright_{ t - k }^{ t } ,  \mathcal{S}'_{ t + 1 } \right> $ (or, equivalently, $ \mathbf{U}\left( \left< \left< \mathcal{O} \upharpoonright_{ 0 }^{ t } , \mathcal{O}'_{ t + 1 } \right> , p \right> \right) = \left< \mathcal{S} \upharpoonright_{ t - k }^{ t } ,  \mathcal{S}'_{ t + 1 } \right> $).
It is immediate to prove that a perfect observation satisfies Definition~\ref{defObservationprinciple} (because the existence of $ p $ directly implies the existence of a constant $ c_\mathcal{O} $ that satisfies Equation~\ref{equationObservationprinciple}) and, indeed, a perfect observation is a particular case of Definition~\ref{defObservationprinciple}.
Thus, an ideal observation takes place when not only the constant $ c_\mathcal{O} $ in Definition~\ref{defObservationprinciple} is small, but also there is a fixed program $ p $ (which does not depend on both the observer and the observed system) that computes the state space trajectory $ \left( \mathcal{S} \upharpoonright_{ t - k }^{ t } ,  \mathcal{S}'_{ t + 1 } \right) $ given the information gathered by the formal observer system.
Thus, a perfect observation is understood to be perfect or ideal not only because the particular observer gathered all the information about the observed system's behaviour so that there is an algorithm that can retrieve the very observed system's behaviour from the internal states of the observer, but also because this holds from the perspective of any possible observer that knows that algorithm.
In other words, a perfect observation is a particular kind of observation in which, although the act of observing itself is subjective with respect to the respective observer, the information gathered by the observer is considered to be objectively sufficient for retrieving the observed system's behaviour by any possible third-party observer.
Changing the scope to stochastic processes instead of deterministic processes, this notion of perfect observation may be tightly connected to the concept of perfect observation of a (stochastically) random variable in \cite{Polani2003}.
\citet{Polani2003} defines a perfect observation when the number of random variables that constitute the observer is sufficiently large so that the conditional entropy of the observed system given these random variables is as small as one wishes.
Then, the increase of intrinsic information within all these variables over time is proposed as a measure of self-organization.
Beforehand, besides the fact that we are dealing with FDDDSs and not stochastic processes, one key distinction of our formalization from the one introduced in \cite{Polani2003} is that we do not assume that the observed system is not perturbed by the act of observing itself---i.e., straightforwardly translating to context of a stochastically generated random variable, we do not assume that the probability distribution of the observed random variable remains unchanged after the act of observing.
Further research is necessary to investigate how the results of the present article are related to emergence and dependency on the observer in the context of stochastic processes.

\section{Emergence of algorithmic information}\label{sectionEmergence}


\subsection{Observer-dependent emergence}\label{sectionWeakemergentalgorithmicinformation}

Intuitively, emergence of algorithmic information occurs when the formal theory professed by the observer is not sufficient for computing, predicting, or completely explaining the object's future behaviour from its constituent parts or prior conditions. 
In the process of trying to explain or predict the behaviour of an observed system (or object), the observer employs the resources available, its own previously held formal knowledge, and the information it could gather from the observation.

The \emph{main idea} of Definition~\ref{defObserverdependentemergence} is that, even if one takes into account equivalent methods to measure the irreducible information content 
(given the presence of the constant $ c_\mathbf{I} $), the error margin of defective information in the observation 
(given the presence of the constant $ c_{ \mathcal{O} } $), and the algorithmic-informational cost of processing all the information that the observer could gather (given the presence of the constant $ c_e $), there is still an insufficient amount of algorithmic information to compute the future behaviour of the observed system.
The presence of these three object-independent constants $ c_\mathbf{I} $, $ c_{ \mathcal{O} } $, and $ c_e $ sets the extent to which the invariance and robustness of the emergence in Definition~\ref{defObserverdependentemergence} hold when the FOS is trying to compute or predict the behaviour of a system.
Because they depend on the FOS and not on the object (i.e., the observed system), they serve the dual purpose of expressing the capabilities of the FOS, while still taking into account the inherent subjectivity of the FOS, which is the distinctive feature of the emergence captured by Definition~\ref{defObserverdependentemergence}.
Note that, contrary to negatively impacting the second type of emergence in Section~\ref{sectionAOIE}, the presence of these three constants in fact emphasizes the strength of the second type of emergence formalized in Definition~\ref{defAOIE}.

%
%
%

In short, extending the same notation we employed in Section~\ref{sectionObservationprinciple}, let $ \left( \mathcal{S} \upharpoonright_{ t - k }^{ t } ,  \mathcal{S}' \upharpoonright_{ t + 1 }^{ t' } \right) $ denote the state space trajectory $   \left( \mathcal{S}_{ t - k } , \dots , \mathcal{S}_{ t } , \mathcal{S}'_{ t + 1 } , \dots , \mathcal{S}'_{ t' } \right) $ and $ \left< \mathcal{S} \upharpoonright_{ t - k }^{ t } ,  \mathcal{S}' \upharpoonright_{ t + 1 }^{ t' } \right> $ denote any arbitrarily chosen encoding of the sequence $ \left( \mathcal{S} \upharpoonright_{ t - k }^{ t } ,  \mathcal{S}' \upharpoonright_{ t + 1 }^{ t' } \right) $.
Also consistently with our notation, we employ $ \left( \mathcal{S} \upharpoonright_{ t }^{ t } \right) $ to denote the single state $  \mathcal{S}_t  $ at time instant $ t $, i.e., $ \left( \mathcal{S} \upharpoonright_{ t }^{ t } \right)  = \mathcal{S}_t $ for every $ t $.

\begin{definition}[Observer-dependent emergence]\label{defObserverdependentemergence}
	Let $ t' \geq t + m $, where $ m \geq 1 $.
	Let $ O $ be the OTM that $ \mathcal{O} $ is simulating.
	Let $ c_e > 0 $ be a constant that may depend on $ \mathcal{O} $, but does not depend on $ \mathcal{S} $. 
	A state space trajectory $  \left( \mathcal{S} \upharpoonright_{ t - k }^{ t } ,  \mathcal{S}' \upharpoonright_{ t + 1 }^{ t' } \right) $ is \emph{emergent with respect to the observer} $ \mathcal{O} $ after the observation of the $ k + 1 $ past states of $ \mathcal{S} $ at time instant $ t $ if
	\begin{equation}\label{equationObserverdependentemergence}
		\mathbf{ I_{ ac } }\left( \left< \mathcal{S} \upharpoonright_{ t - k }^{ t } ,  \mathcal{S}' \upharpoonright_{ t + 1 }^{ t' } \right> \, \middle\vert \, \left< w , \left( \mathcal{O} \upharpoonright_{ 0 }^{ t } ,  \mathcal{O}' \upharpoonright_{ t + 1 }^{ t + m } \right) \right> \right) > c_\mathbf{I} + c_{ \mathcal{O} } + c_e
		\text{ ,}
	\end{equation}
	where $ w $ is the bit string on the first tape of $ O $ that satisfies
	the Definition~\ref{defObservationprinciple} at time instant $ t $.
\end{definition}

The constant $ c_e $ can be arbitrarily large in order to allow high-algorithmic-complexity programs to process prior information and try to compute future behaviour.
However, once its value is fixed, it does not depend on the choice of the observed system (or object).
It can only depend on the observer.

This way, once the three constants $ c_\mathbf{I} $, $ c_{ \mathcal{O} } $, and $ c_e $ are fixed for each formal observer system, the \emph{invariance} of $ \mathbf{ I_{ ac } }\left( \left< \mathcal{S} \upharpoonright_{ t - k }^{ t } ,  \mathcal{S}' \upharpoonright_{ t + 1 }^{ t' } \right> \, \middle\vert \, \left< w , \left( \mathcal{O} \upharpoonright_{ 0 }^{ t } ,  \mathcal{O}' \upharpoonright_{ t + 1 }^{ t + m } \right) \right> \right) $ with respect to the observer follows directly from the invariance mentioned in Section~\ref{sectionAICofOs}.
In the same manner, the \emph{minimality} (\emph{incompressibility} or \emph{irreducibility}) of $ \mathbf{ I_{ ac } }\left( \left< \mathcal{S} \upharpoonright_{ t - k }^{ t } ,  \mathcal{S}' \upharpoonright_{ t + 1 }^{ t' } \right> \, \middle\vert \, \left< w , \left( \mathcal{O} \upharpoonright_{ 0 }^{ t } ,  \mathcal{O}' \upharpoonright_{ t + 1 }^{ t + m } \right) \right> \right) $ with respect to the observer follows directly from the minimality mentioned in Section~\ref{sectionAICofOs}. 
Such invariance and minimality hold because $ \mathbf{ I_{ ac } }\left( \left< \mathcal{S} \upharpoonright_{ t - k }^{ t } ,  \mathcal{S}' \upharpoonright_{ t + 1 }^{ t' } \right> \, \middle\vert \, \left< w , \left( \mathcal{O} \upharpoonright_{ 0 }^{ t } ,  \mathcal{O}' \upharpoonright_{ t + 1 }^{ t + m } \right) \right> \right) $ is an equivalence class that measures (conditional) irreducible information content and that does not vary by the formal method of measuring irreducible information content, universal programming language, or encoding method that the observer chooses.

At first glance, since the constant $ c_e $ can be arbitrarily large, one might think that it is possible to cancel any presence of necessary extra algorithmic information to compute the future behaviour of $ \mathcal{S} $.
This is \emph{not} the case, i.e.,
no matter how large a value one chooses for $ c_e $, there will be FDDDSs for which Equation~\ref{equationObserverdependentemergence} holds.
Suppose that there is $ c_e $ such that, for every $ t' $, there is $ m \leq t' $ such that $ \mathbf{ I_{ ac } }\left( \left< \mathcal{S} \upharpoonright_{ t - k }^{ t } ,  \mathcal{S}' \upharpoonright_{ t + 1 }^{ t' } \right> \, \middle\vert \, \left< w , \left( \mathcal{O} \upharpoonright_{ 0 }^{ t } ,  \mathcal{O}' \upharpoonright_{ t + 1 }^{ t + m } \right) \right> \right) \leq c_\mathbf{I} + c_{ \mathcal{O} } + c_e $ holds.
Hence, we would have it that
\begin{equation}\label{equationExampleofnonODE}
\begin{aligned}
\mathbf{K}\left( \left< \mathcal{S} \upharpoonright_{ t - k }^{ t } ,  \mathcal{S}' \upharpoonright_{ t + 1 }^{ t' } \right> \right) 
& \leq 
\mathbf{K}\left( t - k , t , m , t'  \right) + c_\mathbf{I} + c_{ \mathcal{O} } + c_e + \mathbf{O}\left( 1 \right) 
\leq \\
& \leq
\mathbf{O}\left( \log\left( t - k \right) + \log\left( t' \right) \right)
\text{ .}
\end{aligned}
\end{equation}
Equation~\ref{equationExampleofnonODE} holds because, since $ \mathcal{O} $ is a formal observer system and the observation is a single event at time instant $t $ (where $ t < t + m \leq t'$) that does not depend on the value of $ t' $, the algorithmic complexity necessary to compute $ \left< w , \left( \mathcal{O} \upharpoonright_{ 0 }^{ t } ,  \mathcal{O}' \upharpoonright_{ t + 1 }^{ t + m } \right) \right> $ is upper bounded by $ \mathbf{K}\left( w  \right) + \mathbf{K}\left( \mathcal{O} \upharpoonright_{ 0 }^{ 0 }  \right) + \mathbf{K}\left(  t , m \right) + \mathbf{ O }\left( 1 \right)$ and the basic inequality
$ \mathbf{K}\left( y \right) \leq \mathbf{K}\left( x \right) + \mathbf{K}\left( y \middle\vert x \right) + \mathbf{O}\left( 1 \right) $ holds in algorithmic information theory.
Thus, by choosing a post-observation state space trajectory $  \left(   \mathcal{S}' \upharpoonright_{ t + 2 }^{ t' } \right) $ and sufficiently larger values of $ \epsilon $ such that 
\begin{equation}\label{equationEpsilonlargerthancomplexityoftime}
\epsilon >  \mathbf{O}\left( \log\left( t - k \right) + \log\left( t' \right) \right) \geq \mathbf{K}\left( t - k , t , m , t' \right) + \mathbf{O}\left( 1 \right) 
\end{equation}
and 
$ \mathbf{K}\left( \left< \mathcal{S} \upharpoonright_{ t - k }^{ t } ,  \mathcal{S}' \upharpoonright_{ t + 1 }^{ t' } \right> \right) > \epsilon $ hold,
we guarantee that any constant $ c_e $ will be overcome (and, therefore, satisfying Equation~\ref{equationObserverdependentemergence}) for some large enough time instant $ t' $.

Indeed, one can always construct a state space trajectory of finite states whose global algorithmic information content is larger than $ \epsilon $ for any $ \epsilon $ that satisfies Equation~\ref{equationEpsilonlargerthancomplexityoftime}.
For example, if the state space $ X_\mathcal{S} $ of $ \mathcal{S} $ is composed of all possible $ n $-bit-length strings and $ t - k \ll t' $, then one can construct a sufficiently long state space trajectory $ \left( \mathcal{S} \upharpoonright_{ t - k }^{ t } ,  \mathcal{S}' \upharpoonright_{ t + 1 }^{ t' } \right)  $ in which the ordering of occurrence of the states produces a final string of length $ n\left( t' - t + k + 1  \right) $ that is incompressible, i.e., $ \mathbf{K}\left( \left< \mathcal{S} \upharpoonright_{ t - k }^{ t } ,  \mathcal{S}' \upharpoonright_{ t + 1 }^{ t' } \right> \right) > n\left( t' - t + k + 1  \right) - \mathbf{O}\left( 1 \right) $, which in turn is a quantity that grows much faster than $  \log\left( t - k \right) + \log\left( t' \right) $.

Another way to obtain high-algorithmic-complexity state spaces trajectories is with FDDDSs whose states $ \mathcal{S}_t $ themselves are large enough so that some of them can have a sufficiently larger algorithmic information content.
For example, suppose the state space $ X_\mathcal{S} $ of $ \mathcal{S} $ is composed of all possible $ n $-bit-length strings, where $ n \gg \log\left( n \right) $.
Then, one constructs a state space trajectory $ \left( \mathcal{S} \upharpoonright_{ 0 }^{ n - 1 } ,  \mathcal{S}' \upharpoonright_{ n }^{ n + n } \right)  $ so that $ \left( \mathcal{S} \upharpoonright_{ 0 }^{ n - 1 } , \mathcal{S}' \upharpoonright_{ n }^{ n  } \right) $ is a computable state space trajectory given $ n  $ as input, $ \mathbf{K}\left( \mathcal{S}' \upharpoonright_{ n + 1 }^{ n + 1 } \right) \geq n - \mathbf{O}\left( 1 \right) $, and every state after time instant $ n $ is just a repetition of the state $ \left( \mathcal{S}' \upharpoonright_{ n + 1 }^{ n + 1 } \right) $.
Thus, even though the state space trajectory $ \left( \mathcal{S} \upharpoonright_{ 0 }^{ n - 1 } , \mathcal{S}' \upharpoonright_{ n }^{ n  } \right) $ is computable and the remaining state space trajectory $ \left( \mathcal{S}' \upharpoonright_{ n + 1 }^{ n + n } \right) $ is maximally redundant, we will have it that $ \mathbf{K}\left( \mathcal{S} \upharpoonright_{ 0 }^{ n - 1 } ,  \mathcal{S}' \upharpoonright_{ n }^{ n + n } \right) \geq n - \mathbf{O}\left( 1 \right) $, which is in turn much larger than $  \log\left( 0 \right) + \log\left( n + n \right) =  \log\left( t - k \right) + \log\left( t' \right) $.

In Section~\ref{sectionFOS}, we presented a case in which a formal observer system cannot compute (i.e., prove) that a sentence is true in a particular formal axiomatic theory, but another formal observer system can.
This is a subjective dependency on the observer's prior formal knowledge that also occurs in ODE.
The future behaviour of the observed FDDDS may appear emergent to a formal observer system, while non emergent to another formal observer system.
This is because, if a finite \emph{extra} amount of algorithmic information is sufficient for the first formal observer system to predict the emergent behaviour of the observed FDDDS, then this finite extra amount of algorithmic information can always be converted into a new extended version of the formal theory, which the first observer had.
Hence, the second formal observer system equipped with this new extended formal theory can compute the behaviour of the observed FDDDS that was considered to be emergent to the first observer.
To the second observer, the behaviour of the observed FDDDS ceases to appear emergent.
Thus, in the cases which the emergence of algorithmic information results from a lack of a \emph{finite} amount of algorithmic information, these emergence phenomena can be classified as being dependent on the observer precisely because of this dependency on the prior formal knowledge.

Formally, this means that the fact that $  \left( \mathcal{S} \upharpoonright_{ t - k }^{ t } ,  \mathcal{S}' \upharpoonright_{ t + 1 }^{ t' } \right) $ satisfies Definition~\ref{defObserverdependentemergence} for an observer $ \mathcal{O}_1 $ does \emph{not} imply that it will necessarily satisfy for another observer $ \mathcal{O}_2 $.
In other words, that a sequence $  \left( \mathcal{S} \upharpoonright_{ t - k }^{ t } ,  \mathcal{S}' \upharpoonright_{ t + 1 }^{ t' } \right) $ appears emergent (as in Definition~\ref{defObserverdependentemergence}) to one observer does not imply that it will appear emergent to another observer.
To demonstrate that there is such another observer $ \mathcal{O}_2 $ that can compute $  \left( \mathcal{S} \upharpoonright_{ t - k }^{ t } ,  \mathcal{S}' \upharpoonright_{ t + 1 }^{ t' } \right) $, it suffices to extend the FAT $ \mathbf{F}_1 $ of $ \mathcal{O}_1 $ with an encoded form of the interpretation of the needed $ c_\mathbf{I} + c_{ \mathcal{O} } + c_e + \mathbf{O}( 1 ) $ bits of algorithmic information in the language of $ \mathbf{F}_1 $, resulting in a FAT $ \mathbf{F}_2 $, and to define $ \mathcal{O}_2 $ as simulating the same OTM of $ \mathcal{O}_1 $ but with $ \mathbf{F}_2 $ in its second tape.
That is, (finite) information can always be converted into an extension of a FAT, and then converted into new formal knowledge about what was supposed to be emergent.
Whenever and wherever there is a sufficient finite amount of algorithmic information that can be employed to compute the behaviour of a system, that emergence is in fact dependent on the observer's previous formal knowledge.
Thus, for any emergence that results from a lack of finite algorithmic information---thereby satisfying Definition~\ref{defObserverdependentemergence}--- one in fact has a type of \emph{observer-dependent emergence} (ODE).

Following the equivalence of Turing machines and computable FDDDSs from Lemma~\ref{lemmaFDDDSequivalencewithObjects} as discussed in Section~\ref{sectionAICofCDSs}, Definition~\ref{defObserverdependentemergence} can assume the alternative form by simply replacing Equation~\ref{equationObserverdependentemergence} with
\begin{equation}\label{equationObserverdependentemergenceforTMs}
	\mathbf{ I_{ ac } }\left( \mathbf{U}\left( \left< t' , p \right> \right) \, \middle\vert \, \left< w , t + m , O \right> \right) > c_\mathbf{I} + c_{ \mathcal{O} } + c_e
\end{equation}
in the case where $ p $ is an observed TM, $ O $ is an OTM, and $ w $ is the information received by $ O $ at time instant $ t $.

\subsubsection{Models and examples of observer-dependent emergence}\label{sectionExamplesofODE}


In \cite{Bedau1997}, a weakly emergent phenomenon is defined as one for which the macro-level states of a system can only be derived by simulating the system itself.
Later, \citet{Bedau2010} refines the notion of derivability in this definition, introducing the notion of explanatory incompressibility.
For example, in Conway's Game of Life, one cannot in general decide from the initial configurations whether or not a macrostate behaviour will have a certain property.
Only by simulating the game would it be possible to gain sufficient irreducible information about whether or not the macrostate behaviour has a specific property.
\citet{Bedau1997,Bedau2010} characterises weak emergent in an informal approach, which does not specify the role of the observer, how the micro-level states are observed, how one decides whether or not something is derivable or incompressible, and the time instants that the events are occurring. 
For example, in the case of emergence at a macro-level of a system S from the micro-level D (see also Section~\ref{sectionExamplesofAOIEinholisticsystems} for more discussion on the holistic variant of emergence) \citet{Bedau1997} defines:
\begin{quotation}
	``Macrostate P of S with microdynamic D is weakly emergent iff P can be
derived from D and S’s external conditions but only by simulation.''
\end{quotation}
And in \cite{Bedau2010} it is stated as:
\begin{quotation}
	``If P is a macro property of some system S, then P is weakly emergent if an only if P has a generative explanation from all of S’s prior micro facts, but only in an incompressible way.''
\end{quotation}
For the present purposes, we can assume a free interpretation of the notions raised in \cite{Bedau1997,Bedau2010} and translate both definitions into the context of FDDDSs, algorithmic perturbations, and algorithmic information as slight variation of Definition~\ref{defObserverdependentemergence} by replacing Equation~\ref{equationObserverdependentemergence} and so forth with:
there is a halting program $p$ and there is another FOS $ \mathcal{O}_2 $ that simulates $p$ such that, for every $h$ with $ t - k \leq h \leq t'  $,
\begin{equation}\label{equationSimulationconditionODE}
	 \mathbf{U}\left( \left< h , p \right> \right) =  \left< \mathcal{S} \upharpoonright_{ t - k }^{ t } ,  \mathcal{S}' \upharpoonright_{ t + 1 }^{ h } \right>\text{ ,}
\end{equation} 
and
\begin{equation}\label{equationBedauODE}
	\mathbf{ I_{ ac } }\left( \left< \mathcal{S} \upharpoonright_{ t - k }^{ t } ,  \mathcal{S}' \upharpoonright_{ t + 1 }^{ t' } \right> \, \middle\vert \, \left< w , \left( \mathcal{O} \upharpoonright_{ 0 }^{ t } ,  \mathcal{O}' \upharpoonright_{ t + 1 }^{ t + m } \right) \right> \right) > c_\mathbf{I} + c_{ \mathcal{O} } + c_e
	\text{ ,}
\end{equation}
where $ w $ is the bit string on the first tape of $ O $ that satisfies
the Definition~\ref{defObservationprinciple} at time instant $ t $.

In other words, besides the usual incompressibility condition stated in Equation~\ref{equationObserverdependentemergence}, we just added to Definition~\ref{defObserverdependentemergence} the condition of the existence of another observer that can compute the state space trajectory by simulating the very state space trajectory.
This way, one can easily demonstrate that such a ``simulation irreducibility via explanatory incompressibility'' in \cite{Bedau1997,Bedau2010} is implied by our Definition~\ref{defObserverdependentemergence}.
To this end, suppose there is a way for the FOS $ \mathcal{O} $ to compute the state space trajectory $  \left( \mathcal{S} \upharpoonright_{ t - k }^{ t } ,  \mathcal{S}' \upharpoonright_{ t + 1 }^{ t' } \right) $ of a FDDDS $ \mathcal{S} $ in the $ t + m $ time steps of $ \mathcal{O} $ without simulating the FDDDS.
Therefore, since $ \mathcal{O} $ is a FOS that simulates the observer Turing machine $ O $, we would have it that there is a sufficiently large $ c_e $ such that
\begin{equation*}
	\mathbf{ I_{ ac } }\left( \left< \mathcal{S} \upharpoonright_{ t - k }^{ t } ,  \mathcal{S}' \upharpoonright_{ t + 1 }^{ t' } \right> \, \middle\vert \, \left< w  , \left( \mathcal{O} \upharpoonright_{ 0 }^{ t } ,  \mathcal{O}' \upharpoonright_{ t + 1 }^{ t + m } \right) \right> \right) \leq c_\mathbf{I} + c_{ \mathcal{O} } + c_e 
	\text{ ,} 
\end{equation*}
which directly contradicts Definition~\ref{defObserverdependentemergence}.
That is, for sufficiently small $ c_e $ in comparison to $ \mathbf{K}\left( \left< \mathcal{S} \upharpoonright_{ t - k }^{ t } ,  \mathcal{S}' \upharpoonright_{ t + 1 }^{ t' } \right> \right) $, the ODE in Definition~\ref{defObserverdependentemergence} implies the weak emergence in \cite{Bedau1997,Bedau2010}.
Conversely, since our interpretation of the ``simulation irreducibility via explanatory incompressibility'' only adds one condition to Definition~\ref{defObserverdependentemergence} (see Equation~\ref{equationSimulationconditionODE}), it is straightforward to show that the weak emergence in \cite{Bedau1997,Bedau2010} implies that there are $ \mathcal{O} $, $ c_e $ and $ t' > t + m  $ for which Definition~\ref{defObserverdependentemergence} is satisfied.


It is claimed in \cite{Bedau1997} that the simulation irreducibility is a property that is not dependent on the current limited knowledge of the observer:
\begin{quotation}
	``For P to be weakly emergent, what matters is that there is a derivation of P from
D and S’s external conditions and any such derivation is a simulation. 
	It does not
matter whether anyone has discovered such a derivation or even suspects that it
exists. 
	If P is a weakly emergent, it is constituted by, and generated from, the
system’s underlying microdynamic, whether or not we know anything about this.
	Our need to use a simulation is due neither to the current contingent state of our
knowledge nor to some specifically human limitation or frailty. 
	Although a Laplacian supercalculator would have a decisive advantage over us in simulation speed, she would still need to simulate. Underivability without simulation is a purely formal notion concerning the existence and nonexistence of certain kinds of derivations of macrostates from a system’s underlying dynamic.''
\end{quotation}
Under our interpretation of the ``simulation irreducibility via explanatory incompressibility'', this claim becomes true or not depending on the whether or not one restricts the possibilities of the formal theory that any FOS might have access to.
For example, if every observer (which does not simulate the observed system) can only know the same FAT and all of them are subjected to the same constants $ c_e $, $ c_\mathcal{O} $ and $ c_\mathbf{I} $, then one can argue that the claim holds indeed.
However, in case any formal theory may be encoded into the second tape of a FOS, we can now employ Definition~\ref{defObserverdependentemergence} to demonstrate that this claim becomes false.
Let $ \mathcal{O}_1 $ and $ \mathcal{S} $ be FDDDSs that, for infinitely many $ p $, there are $ t $ and $ t' $ with
\begin{equation}\label{equationNonsimulationcondition}
	t' \geq \mathbf{K}\left( t' \right) > \left( t + m \right)^3 + \mathbf{K}\left( \left< w , \left( \mathcal{O}_1 \upharpoonright_{ 0 }^{ t } ,  \mathcal{O}'_1 \upharpoonright_{ t + 1 }^{ t + m } \right) \right> \right) + c_\mathbf{I} + c_{ \mathcal{O} } + \left| p \right| + \mathbf{O}(1)
\end{equation}
and $ m > 1 $ such that  
\begin{equation*}\label{}
\mathbf{ I_{ ac } }\left( 
\left< \mathcal{S} \upharpoonright_{ t - k }^{ t } ,  \mathcal{S}' \upharpoonright_{ t + 1 }^{ t' } \right> \, 
\middle\vert \, 
\left< w , \left( \mathcal{O}_1 \upharpoonright_{ 0 }^{ t } ,  \mathcal{O}'_1 \upharpoonright_{ t + 1 }^{ t + m } \right) \right> 
\right) 
> 
c_\mathbf{I} + c_{ \mathcal{O} } + \left| p \right| + \mathbf{O}(1)
\text{ .}
\end{equation*}
Clearly, $ \mathcal{O}_1 $ cannot always predict or compute the state space trajectory $   \left( \mathcal{S'} \upharpoonright_{ t + 1 }^{ t' } \right) $ during the $ t + m $ time steps of $ \mathcal{O}_1 $.
However, for any $ p , t $, and $ t' $ for which this holds, we have already shown in Section~\ref{sectionWeakemergentalgorithmicinformation} how one can construct another $ \mathcal{O}_2 $ that can compute $ \left( \mathcal{S} \upharpoonright_{ t - k }^{ t } ,  \mathcal{S}' \upharpoonright_{ t + 1 }^{ t' } \right) $ in $ t + m $ time steps.
In addition, since Equation~\ref{equationNonsimulationcondition} holds, then neither $ \mathcal{O}_1 $ nor $ \mathcal{O}_2 $ can be simulating $ \left( \mathcal{S} \upharpoonright_{ t - k }^{ t } ,  \mathcal{S}' \upharpoonright_{ t + 1 }^{ t' } \right) $ during the $ t + m $ time steps in the first place.
This observer dependency also trickles down to the resource-bounded case.
For example, assuming \textbf{P-TIME} $ \neq $ \textbf{NP-TIME}, one can construct a $ \left( \mathcal{S} \upharpoonright_{ t - k }^{ t } ,  \mathcal{S}' \upharpoonright_{ t + 1 }^{ t' } \right) $, which is equivalent to a problem in \textbf{NP-TIME}, such that a polynomial-time $ \mathcal{O}_1 $ cannot compute it, while another polynomial-time $ \mathcal{O}_2 $ that has a polynomial-length succinct certificate encoded in its second tape can.

Thus, since the ``simulation irreducibility via explanatory incompressibility'' in \cite{Bedau1997,Bedau2010} implies that there are conditions in which Definition~\ref{defObserverdependentemergence} is satisfied, and vice-versa, we argue that the weak emergence described in \cite{Bedau1997,Bedau2010} can be understood as an informal alternative, but one that is \emph{equivalent} to the ODE captured by Definition~\ref{defObserverdependentemergence}.

Another model for FDDDSs in which a type of emergence occurs may be found in \cite{Adams2017,Adams2017a}.
If the interaction of a (finite) dynamical system $ A $ with its environment $ \mathcal{E} $ (i.e., another finite dynamical system) gives rise to a recurrent state space trajectory whose length is greater than the length of all the other recurrent state space trajectories of any isolated system of the same size as $ A $, the pair of systems $ A $ and $ \mathcal{E} $ is said to exhibit \emph{unbounded evolution} (UE).
If the newly emerging recurrent state space trajectory from the interaction between $ A $ and $ \mathcal{E} $ is \emph{not} contained in any of the other recurrent state space trajectories of any isolated system with the same size as $ A $, the pair of systems $ A $ and $ \mathcal{E} $ is said to exhibit \emph{innovation} (INN). 
In addition, since interaction with the environment can introduce state-dependent changes in the evolution rules of the system $ A $, the increase of the recurrence time shown in the models investigated in \cite{Adams2017} can be classified as an example of emergence through downward (or top-down) causation \cite{Adams2017,Walker2012,Davies2008}. 

Note that, in the models of cellular automata (CA) in \cite{Adams2017}, the interaction with the environment can produce changes, i.e., perturbations, in the e.r. of system $ A $.
Hence it differs from the AP in Definition~\ref{defAlgorithmicperturbation} because the latter impacts the states of the system $ A $ instead of its e.r..
In fact, one can reduce each state-dependent rule perturbation on a CA $ A $ in \cite{Adams2017} to an equivalent AP on an equivalent universal CA that emulates $ A $.
To this end, simply note that every finite CA is computable by a TM, and that there are universal cellular automata (for example, ECA Rule 110) that can simulate any Turing machine. 
Thus, for each state-dependent rule perturbation in \cite{Adams2017}, one constructs the equivalent AP in the same manner as in the proof of Lemma~\ref{lemmaSequence computabilityofO}.

Hence, in order to show that Definition~\ref{defObserverdependentemergence} implies UE and INN, it suffices to choose a large enough constant $ c_e $ that depends on the observer $ \mathcal{O} $ but does not depend on $ \mathcal{S} $ nor on $ \mathcal{E} $, where both $ \mathcal{S} $ and $ \mathcal{E} $ are FDDDSs. 
Let $ n_e = \max\left\{ \left| \mathcal{E}_t \right| \, \middle\vert t \in T \right\} $ be the size of $ \mathcal{E} $ and $ n_s = \max\left\{ \left| \mathcal{S}_t \right| \, \middle\vert t \in T \right\} $ be the size of $ \mathcal{S} $.
In the particular case of ECAs, then $ n_e $ and $ n_s $ become the respective widths of each ECA.
Let $\left| X_{ \mathcal{S} } \right| $ be the maximum number of distinct states of a system $ \mathcal{S} $ and $ \left| X_{ \mathcal{E} } \right| $ be the maximum number of distinct states of the environment $ \mathcal{E} $.
In the case of ECAs, we would have $\left| X_{ \mathcal{S} } \right| = 2^{ n_s } $ and $ \left| X_{ \mathcal{E} } \right| = 2^{ n_e } $.
Thus, the maximum number of possible state space trajectories 
is upper bounded by $ {\left| X_{ \mathcal{S} } \right| }^{ t_r }  $, where
$ t_r $ is the recurrence time of $ \mathcal{S} $ when it is interacting with $ \mathcal{E} $.
Now, let $ t_p $ be the largest recurrence time of any isolated $ \mathcal{S} $. 
We choose a constant  
\begin{equation}\label{equationODEimpliesUE}
	c_e > \left| p' \right| + \mathbf{O}\left( \log_2\left( {\left| X_{ \mathcal{S} } \right| }^{ t_p } \right) \right)
	+ \mathbf{O}\left( \log_2\left( t_p \right) \right)
	+ \mathbf{O}\left( \log_2\left( t_r \right) \right)
	\text{ ,}
\end{equation}
where $ p' $ is the program that 
reads the index of a recurrent state space trajectory of length $ t'' \leq t_p $ (index which is encoded in $  \mathbf{O}\left( \log_2\left( {\left| X_{ \mathcal{S} } \right| }^{ t_p  } \right) \right) $ bits) and returns a state space trajectory by extending and repeating the $ t'' $ update steps (where $ t'' $ is an integer encoded in $  \mathbf{O}\left( \log_2\left( t_p \right) \right) $ bits) until the $ t_r $-length sequence of states is completed (where $ t_r $ is an integer encoded in $  \mathbf{O}\left( \log_2\left( t_r \right) \right) $ bits).
Clearly, once $ t_r $ can be arbitrarily larger than $ t_p $, if 
a state space trajectory $ \left( \mathcal{S} \upharpoonright_{ t - k }^{ t } ,  \mathcal{S}' \upharpoonright_{ t + 1 }^{ t' } \right) $
satisfies Definition~\ref{defObserverdependentemergence} with such a constant $ c_e $ (where $ \mathcal{S} \upharpoonright_{ t - k }^{ t } $ corresponds to $ \mathcal{S} $ isolated, $ \mathcal{S}' \upharpoonright_{ t + 1 }^{ t' } $ corresponds to $ \mathcal{S} $ interacting with $ \mathcal{E} $, $ t = t_p $, $ k = t $, $ t' = t_r + t_p $, and
$ t $ is the time instant at which the observation of the isolated $ \mathcal{S} $ occurred), then $ \left( \mathcal{S}' \upharpoonright_{ t + 1 }^{ t' } \right) $ cannot be any recurrent state space trajectory of length $ \leq t_p $, and
therefore this state space trajectory satisfies both the definitions of UE and INN in \cite{Adams2017}.

Also note that a slight variation of such a program $ p' $ can be employed to prove that UE and INN define a type of emergence that is dependent on the observer's prior knowledge.
To this end, simply note that, for any recurrent state space trajectory of length $ t_r $, there is a constant $ c_e \leq \left| p' \right| + \mathbf{O}\left( \log_2\left( {\left| X_{ \mathcal{S} } \right| }^{ t_r } \right) \right)
+ \mathbf{O}\left( \log_2\left( t_r \right) \right) $ that negates Equation~\ref{equationObserverdependentemergence}.

Thus, the kind of emergence from UE and INN is dependent on the observer's formal knowledge and there is at least one constant $ c_e $ for which UE and INN is implied by the ODE in Definition~\ref{defObserverdependentemergence}.
In attempting to show that the two approaches are equivalent, it is important to remark that models displaying an empirical tendency of increasing algorithmic complexity were investigated in \cite{Adams2017}.
In this line of inquiry, the inverse problem (that is, to prove that a FDDDS displaying UE and INN \emph{always} implies that there is at least one constant $ c_e $ for which the state space trajectory of Definition~\ref{defObserverdependentemergence} is satisfied) is necessary theoretical research that remains to be done.

%



\subsection{Asymptotically observer-independent emergence}\label{sectionAOIE}

The ODE studied in Section~\ref{sectionWeakemergentalgorithmicinformation} establishes a formal mathematical characterisation of emergence that takes into account observation errors or distortions, the arbitrarily chosen mathematical method of measuring information content, and any information cost to process prior knowledge and the information gathered in the act of observing.
The next question that naturally arises is whether such an approach can be extended to formalise an emergent phenomenon that continues to be emergent for any observer.
In this regard, the present section tackles the problem by presenting a precise definition and demonstrating that \emph{asymptotically observer-independent emergence} (AOIE) of algorithmic information exists in mathematical models.

First, we naturally extend our notation so 
we can further explore the necessity of extra algorithmic information when a sufficiently large amount of time steps has passed.
In the general sense, a relational structure $ A $ is said to be an \emph{extension} of another relational structure $ B $ iff not only the domain of $ A $, but also all the relations that define $ A $ are included into the relational structure $ B $ \cite{Hodges1993}.
Let $ \mathbf{S} = \left(  \mathcal{S}^{ ( 1 ) } , \mathcal{S}^{ ( 2 ) }, \dots , \mathcal{S}^{ ( k ) } , \dots \right) $ denote a \emph{discrete deterministic dynamical system} that is composed of a sequence of nested FDDDSs, where:
the FDDDS $  \mathcal{S}^{ ( i + 1 ) } $ is an extension of the FDDDS $  \mathcal{S}^{ ( i ) } $ for every $ i > 0 $;
and the transition from $  \mathcal{S}^{ ( i ) } $ to $  \mathcal{S}^{ ( i + 1 ) } $ occurs at time instant $ t_{ i + 1 } - 1 $.
Thus, by translating the definition of extension into the context of FDDDS, this is equivalent to say that the evolution rule $ f_\mathbf{S} $ of $ \mathbf{S} $ is such that, for every $ i > 0 $ with $ t_{ i + 1 } - 1 >  t \geq t_i  $, one has 
\begin{equation*}
f_\mathbf{S}\left( \left(\mathcal{S}^{ ( i ) } \upharpoonright_{ t }^{ t }\right) , \left(e_{\mathcal{S}^{ ( i ) }} \upharpoonright_{ t }^{ t }\right) , t  \right) 
= 
f_{ \mathcal{S}^{ ( i ) } }\left( \left(\mathcal{S}^{ ( i ) } \upharpoonright_{ t }^{ t }\right) , \left(e_{\mathcal{S}^{ ( i ) }} \upharpoonright_{ t }^{ t }\right) , t  \right)
\end{equation*} 
and 
\begin{equation*}
f_\mathbf{S}\left( \left(\mathcal{S}^{ ( i ) } \upharpoonright_{ t_{ i + 1 } - 1 }^{ t_{ i + 1 } - 1 }\right) , \left(e_{\mathcal{S}^{ ( i ) }} \upharpoonright_{ t_{ i + 1 } - 1 }^{ t_{ i + 1 } - 1 }\right)  , t_{ i + 1 } - 1  \right) 
= 
\left( \mathcal{S}^{ ( i + 1 ) } \upharpoonright_{ t_{ i + 1 } }^{ t_{ i + 1 } } \right)
\text{ .}
\end{equation*}
In other words, $ \mathcal{S}^{ ( 2 ) } $ extends $ \mathcal{S}^{ ( 1 ) } $ because the FDDDS $ \mathcal{S}^{ ( 2 ) } $ behaves in the exact same way as the FDDDS $ \mathcal{S}^{ ( 1 ) } $ until time instant $ t_{ 2 } - 1 $, $ \mathcal{S}^{ ( 3 ) } $ extends $ \mathcal{S}^{ ( 2 ) } $ because $ \mathcal{S}^{ ( 3 ) } $ behaves in the exact same way as $ \mathcal{S}^{ ( 2 ) } $ until time instant $ t_{ 3 } - 1 $, and so on.
The sequence of all these extensions constitutes the FDDDS $ \mathbf{S} $.
Consistently with the notation, notice that a FDDDS $ \mathcal{S} $ is a special case of $ \mathbf{S} $ in which $ \mathbf{S} = \left( \mathcal{S} , \mathcal{S} , \mathcal{S} , \dots \right) $.

Now, we formalise emergence of algorithmic information that is able to eventually surpass any previous formal knowledge that the observer held or may come up with.
The main idea of Definition~\ref{defAOIE} is that, for any  
equivalent way to measure the irreducible information content (given the presence of the constant $ c_\mathbf{I} $), the error margin of defective information in the observation (given the presence of the constant $ c_{ \mathcal{O} } $), and the algorithmic-informational cost of processing all the previous information that the observer has (given the presence of the constant $ c_e $), there is a certain stage of the system $ \mathbf{S} $ where it begins to lack the amount of algorithmic information necessary to compute the future behaviour of the observed system. 
In other words, for every FOS, there is a certain stage at which the subsequent behaviour of $ \mathbf{S} $ begins to display ODE (as in Definition~\ref{defObserverdependentemergence}).

\begin{definition}[Asymptotically observer-independent emergence]\label{defAOIE}
	Let $ \mathbf{S} $ be an arbitrary discrete deterministic dynamical system well defined for every time instant (i.e., $ \left| T \right| = \infty $).
	A system $ \mathbf{S} $ is \emph{asymptotically observer-independently emergent} if, for every $ \mathcal{O} $, for every time instant $ t $, and for every $ c_\mathbf{I} , c_{ \mathcal{O} } , c_e > 0 $, there is $ t_e \in T $ such that, for every $ t' \geq t_e $, one has it that 
	\begin{equation}\label{equationAOIE}
	\mathbf{ I_{ ac } }\left( \left< \mathbf{S} \upharpoonright_{ t - k }^{ t } ,  \mathbf{S}' \upharpoonright_{ t + 1 }^{ t' } \right> \, \middle\vert \, \left< w , \left( \mathcal{O} \upharpoonright_{ 0 }^{ t } ,  \mathcal{O}' \upharpoonright_{ t + 1 }^{ t + m } \right) \right> \right) > c_\mathbf{I} + c_{ \mathcal{O} } + c_e
	\text{ ,}
	\end{equation}
	where $ t' \geq t + m $, $ m \geq 1 $, and $ w $ is the bit string in the first tape of $ O $ that satisfies the Definition~\ref{defObservationprinciple} at time instant $ t $.
\end{definition}

Clearly, the AOIE in Definition~\ref{defAOIE} inherits the \emph{invariance} and \emph{minimality} of the ODE in Definition~\ref{defObserverdependentemergence}.
The distinctive aspect now is the fact that although there might be a FOS that can compute a finite-length state space trajectory of $ \mathbf{S} $, the AOIE guarantees that this will eventually cease to happen.
For this reason, the emergence of algorithmic information in Definition~\ref{defAOIE} is guaranteed to be \emph{independent} of the observer, but only at the \emph{asymptotic limit}.
The existence of the time instant $ t_e $ ensures that there is a phase transition for which, if the behaviour of $ \mathbf{S} $ is not emergent to a certain FOS, then it will start to be emergent after time instant $ t_e $.
Indeed, at each stage of $ \mathbf{S} $, since it is a FDDDS, the strongest emergence that $ \mathbf{S} $ can display is the ODE in Definition~\ref{defObserverdependentemergence}, but eventually any other FOS that might try to compute the behaviour of $ \mathbf{S} $ will be outdone after time instant $ t_e $. 
We shall demonstrate in the next Sections~\ref{sectionExamplesofAOIEinevolutionarysystems} and \ref{sectionExamplesofAOIEinholisticsystems} that there are indeed abstract mathematical models that satisfy Definition~\ref{defAOIE}.

When dealing with programs or Turing machines instead of dynamical systems, Definition~\ref{defAOIE} can assume the alternative and more simplistic form.
We say an infinite collection $ \mathbf{P} $ of programs displays \emph{asymptotically observer-independent emergence} if, for every $ O $, for every time instant $ t $, and for every $ c_\mathbf{I} , c_{ \mathcal{O} } , c_e > 0 $, there are $ p \in \mathbf{P} $ and $ t_e \in T $ such that, for every $ t' \geq t_e $, one has it that 
\begin{equation}\label{equationAOIEforTMs}
\mathbf{ I_{ ac } }\left( \mathbf{U}\left( \left< t' , p \right> \right) \, \middle\vert \, \left< w , t + m , O \right> \right) > c_\mathbf{I} + c_{ \mathcal{O} } + c_e
\text{ ,}
\end{equation}
where $ t' \geq t + m $, $ m \geq 1 $, and $ w $ is the bit string in the first tape of $ O $ that satisfies the Definition~\ref{defObservationprinciple} at time instant $ t $.

\subsubsection{A model and example in evolutionary systems}\label{sectionExamplesofAOIEinevolutionarysystems}

In \cite{Chaitin2012,Chaitin2013,Hernandez-Orozco2018}, a theoretical model for the open-ended evolution of programs (or Turing machines) is presented with the purpose of obtaining a mathematical proof of Darwinian evolution within the framework of algorithmic information theory (AIT).
Inspired by (but not limited to) evolutionary biology, this field is called \emph{metabiology} and in a general sense it constitutes a pursuit of mathematical proofs of meta-level fundamental properties and quintessential ``laws'' in evolutionary systems \cite{Chaitin2018}.

The cumulative evolution model is defined in \cite{Chaitin2012,Chaitin2013} as a sequence of sole (resource-unbounded) Turing machines that evolve over time due to the transformations effected by randomly generated algorithmic mutations:
one starts with an arbitrary TM and then subsequently applies randomly generated algorithmic mutations so that, if a mutation leads to a fitter TM than the previous TM, the mutant TM replaces the old one, and so on.
Thus, in accordance with evolutionary biology, not only these Turing machines are subjected to randomly generated mutations and natural selection, but also may inherit information from their predecessors.
\citet{Chaitin2012} presents a theoretical analysis of the expected sufficient number of algorithmic mutations for reaching a fitness value that necessarily requires $n$ bits of irreducible information content to be computed.
Indeed, in the cumulative evolution model, $n$ bits of algorithmic information is proved to be achieved in a realistically small number of randomly generated mutations. 
Due to the known mathematical properties of algorithmic information such as invariance and minimality (see Section~\ref{sectionAICofOs}), this shows that a quantity of irreducible information content is achieved in a realistically fast mutation time through randomly generated mutations applied to evolving programs that can inherit past information.
In particular, \citet{Chaitin2012} demonstrates that $ n $ bits of algorithmic complexity is expected to be reached after $ \mathbf{ O }( n^2 ( \log(n) )^2 ) $ randomly generated algorithmic mutations.
This result is achieved by employing a theoretical analysis of the resulting algorithmic complexity from certain algorithmic mutations that are expected to occur over time.
\citet{Abrahao2015,Abrahao2016} then demonstrated that the results for resource-\emph{unbounded} Turing machines in the former cumulative evolution model trickles down to the realistic resource-\emph{bounded} case: 
$ n $ bits of \emph{time-bounded} algorithmic complexity is expected to be reached after $ \mathbf{ O }( n^2 ( \log(n) )^2 ) $ randomly generated \emph{time-bounded} algorithmic mutations in the cumulative evolution of \emph{time-bounded} Turing machines.

These abstract evolutionary models were then corroborated by empirical results
in \cite{Hernandez-Orozco2018a}, not only showing that randomly generated algorithmic mutations produce a speed-up in adaptation in comparison to the uniformly random point mutations (which are the usual random mutations under consideration in mainstream models based on evolutionary modern synthesis), but also that it may be related to explanations of the occurrence of modularity, diversity explosions, and massive extinctions.

We remark that algorithmic mutations as in \cite{Chaitin2012,Chaitin2013} are exactly the APs we defined in Section~\ref{sectionAlgorithmicperturbations}, except that in these particular evolutionary models, these APs are randomly generated following the usual i.i.d. probability distribution of prefix-free binary sequences.

We can now apply the result from \cite{Chaitin2012} in order to demonstrate the existence of a system that displays \emph{expected} AOIE.
The main idea behind Theorem~\ref{thmAOIEinevolution} is that the cumulative evolution of sole TMs under successive perturbations performed by the randomly generated algorithmic mutations is able to guarantee (with a probability as high as one desires) that the emergence of algorithmic information is larger than any FOS can keep up with in the long run.

\begin{theorem}\label{thmAOIEinevolution}
	Let $ \mathbf{P} $ be a sequence of TMs that results from the cumulative evolution model in \cite{Chaitin2012} after $ t $ successive randomly generated algorithmic mutations.
	Then, $ \mathbf{P} $ satisfies Definition~\ref{defAOIE} with probability arbitrarily close to $1$ as $ t \to \infty $.
	
	\begin{proof}
		
		From \cite{Chaitin2012}, we know that for any initial TM $ P_0 $ and sufficiently large $ t $, the algorithmic complexity of the output of the $ t $-th TM $ P_t $ in the sequence $ \mathbf{P} $ is expected to be at least as large as $  \sqrt[3]{ t }  $, where $ t $ is the number of successive randomly generated algorithmic mutations.
		That is, $ \mathbf{K}\left( \mathbf{U}\left( \left< t , P_t \right> \right) \right) \geq {\boldsymbol \Omega}\left( \sqrt[3]{ t } \right) $ holds with probability arbitrarily close to $ 1 $ as $ t \to \infty $, where $ {\boldsymbol \Omega}\left( \cdot \right) $ is the usual Big-\textbf{Omega}\footnote{ Do not confuse this notation with the halting probability, which is a real number between $ 0 $ and $ 1 $.} notation for asymptotic lower bounds.
		Suppose that there is at least one OTM $ O $ and there are $ c_\mathbf{I} , c_{ \mathcal{O} } $, and $ c_e  $ such that, for every $ P_{ t' } \in \mathbf{P} $ and $ t_e \in T $ with $ t' \geq t_e $, one has it that 
		$
		\mathbf{ I_{ ac } }\left( \mathbf{U}\left( \left< t' , P_{ t' } \right> \right) \, \middle\vert \, \left< w , t_o + m , O \right> \right) \leq c_\mathbf{I} + c_{ \mathcal{O} } + c_e
		\text{ ,}
		$
		where $ t' \geq t_o + m $ and $ w $ is the bit string in the first tape of $ O $ that satisfies the Definition~\ref{defObservationprinciple} at an arbitrary  time instant $ t_o $.
		Then, from basic inequalities in AIT and from our definition of $ \mathbf{ I_{ ac } } $ in Section~\ref{sectionAICofOs}, we would have it that
		\begin{equation*}
			\begin{aligned}
				\mathbf{K}\left( \mathbf{U}\left( \left< t' , P_{ t' } \right> \right) \right)
				& \leq 
				\mathbf{K}\left( t_o +  m \right)
				+ \mathbf{O}(1)
				+ 2 \, c_\mathbf{I} + c_{ \mathcal{O} } + c_e \\
				& \leq 
				\mathbf{O}\left( \log\left( t' \right) \right)
				+ \mathbf{O}(1)
				+ 2 \, c_\mathbf{I} + c_{ \mathcal{O} } + c_e
				\text{ .}
			\end{aligned}
		\end{equation*}
		Therefore, for infinitely many $ t_e $, we would have it that $ \mathbf{K}\left( \mathbf{U}\left( \left< t_e , P_{ t_e } \right> \right) \right) \leq \mathbf{O}\left( \log\left( t_e \right) \right) $ and $ \mathbf{K}\left( \mathbf{U}\left( \left< t_e , P_{ t_e } \right> \right) \right) \geq {\boldsymbol \Omega}\left( \sqrt[3]{ t_e } \right) $, which leads to a contradiction.
	\end{proof}
\end{theorem}

The result in Theorem~\ref{thmAOIEinevolution} is stated using the alternative form of Definition~\ref{defAOIE} for programs (or TMs) so it can easily be translated
to the cumulative evolution model in \cite{Chaitin2012}.
However, it is easy see how Theorem~\ref{thmAOIEinevolution} can be translated into a system $ \mathbf{S} $ instead of a sequence $ \mathbf{P} $.
To this end, replace each program/organism $ P_t $ in \cite{Chaitin2012} and in the proof of Theorem~\ref{thmAOIEinevolution} with the Turing equivalent FDDDS $ \mathcal{S} $, as done in Lemma~\ref{lemmaFDDDSequivalencewithObjects} in Section~\ref{sectionAICofCDSs}.
Then, to achieve the contradiction obtained in the proof of Theorem~\ref{thmAOIEinevolution}, we use the fact that Lemma~\ref{lemmaUncomputabilityofS} is eventually satisfied by those FDDDSs in the sequence $ \mathbf{S} $ composed of sole FDDDSs that evolve over time due to the transformations effected by randomly generated APs.

Emergent phenomena usually can be divided into two kinds \cite{Bar-Yam2004,Chalmers2008}:
a temporal version in which emergence occurs over time, as the system interacts with the environment (for this reason it is called \emph{diachronic} emergence \cite{O'Connor2020,Chalmers2008});
and a holistic (or synchronic) version in which emergence occurs as a distinctive feature of the ``whole'' relative to the parts \cite{O'Connor2020,Chalmers2008}.

Thus, the kind of emergence shown in such an evolutionary AOIE falls under the diachronic variant of AOIE.
In particular, the \emph{open-endedness} proved in \cite{Chaitin2012} strictly refers to the unbounded increase of complexity over time, as the evolution unfolds.
For this reason, it is called \emph{evolutionary open-endedness} \cite{Abrahao2018,Abrahao2019}.
Another example of diachronic emergence and open-endedness is the one presented in \cite{Adams2017}, which we discussed in Section~\ref{sectionExamplesofODE}.
As we have already demonstrated, the distinctive feature of the emergence in the diachronic AOIE is that it is asymptotically independent of the observer's formal knowledge, while the emergence in \cite{Adams2017} is dependent on the observer's formal knowledge.
In this sense, we can also adopt the convention of classifying evolutionary AOIE as \emph{asymptotically observer-independent diachronic open-endedness} and the emergence in \cite{Adams2017} as \emph{observer-dependent diachronic open-endedness}.


\subsubsection{A model and example in networked systems}\label{sectionExamplesofAOIEinholisticsystems}

The pervasiveness of non-homogeneous network topological properties has fostered the recent field of network science and showed its important role in complex systems science \cite{Barabasi2016}.
In this regard, motivated by the pursuit of a unified theory of complexity in network science and complex systems science \cite{Barabasi2009,Michail2018}, the theory of \emph{algorithmic networks} \cite{Abrahao2016b,Abrahao2019,Abrahao2018} allows the investigation of how network topological properties can trigger emergent behaviour that is capable of irreducibly increasing the computational power of the whole network.
An algorithmic (complex) network $ \mathfrak{N} $ is a population of computable systems whose members can share information with each other according to a complex network topology.
Each node of the network is a computable system and each (multidimensional) edge of the network is a communication channel.

In \cite{Abrahao2019}, it is shown that there are network topological properties, such as the small diameter, associated with a computationally cheap and simple communication protocol of plain diffusion that asymptotically trigger an unlimited increase of expected emergent algorithmic complexity of a networked node's final output as the number of nodes increases indefinitely.
The diameter of a network (or graph) is the length of the longest shortest path from any node to any other node \cite{Diestel2017}.
The diameter is said to be small if the diameter grows in a logarithmic order of the number of nodes \cite{Barabasi2016} and
such a small-diameter phenomenon is one of the important properties found in both real-world and synthetic networks \cite{Barabasi2016,Lewis2009}. 

This unlimited increase of expected emergent algorithmic complexity is referred to as \emph{expected emergent open-endedness} (EEOE) \cite{Abrahao2019,Abrahao2018} and---by simplifying the notation from \cite{Abrahao2019,Abrahao2018} to serve our present purposes---it is mathematically defined by
\begin{equation}\label{equationEEOE}
	\lim_{ N \to \infty } \mathbf{E}_{ \mathfrak{N} } 
	\left(
	{ {\displaystyle{\myDelta_{iso}^{net}} \mathbf{ K }} (o_i,c)} 
	\right)
	=
	\infty
	\text{ ,}
\end{equation}
where: 
$ \mathbf{E}_{ \mathfrak{N} }\left( \cdot \right) $ gives the average value over all possible randomly generated nodes in the algorithmic network $ \mathfrak{N} $; $ { {\displaystyle{\myDelta_{iso}^{net}} \mathbf{ K }} (o_i,c)}  $ is the \emph{emergent algorithmic complexity (EAC)} of a node $o_i$ in $c$ communication rounds;
and $ N $ is the total number of nodes in the algorithmic network $ \mathfrak{N} $.
\begin{equation*}
	 { {\displaystyle{\myDelta_{iso}^{net}} \mathbf{ K }} (o_i,c)} = 
	\mathbf{ K }\left( P_{net}\left( o_i , c \right) \right) 
	- 
	\mathbf{ K  }\left( P_{iso}\left( o_i , c \right) \right) 
\end{equation*} 
denotes the difference between the algorithmic complexity of the node $ o_i $ in $ c $ communication rounds when running \emph{networked} and the algorithmic complexity of the node $ o_i $ in $ c $ communication rounds when running \emph{isolated} from any other node, respectively.
More formally:
$ P_{net}\left( o_i , c \right) $ denotes the program that computes the sequence of all the outputs that are sent to $ o_i $'s neighbours in all the communication rounds until $ c $ communication rounds have transpired; 
and $ P_{iso}\left( o_i , c \right) $ denotes the program that computes all the  computation steps of the isolated program $ o_i $ during the $ c $ communication rounds.
Note that the limit of the EEOE eventually neutralises any pair of constants $ c_\mathbf{ I } $ that one may subtract or add to this difference.
Thus, one can equivalently define the EEOE as 
\begin{equation*}\label{equationEEOEforaic}
	\lim_{ N \to \infty } \mathbf{E}_{ \mathfrak{N} } 
	\left(
	{ {\displaystyle{\myDelta_{iso}^{net}} \mathbf{ I_{ ac } }} (o_i,c)} 
	\right)
	=
	\infty
	\text{ .}
\end{equation*}
In case an algorithmic network displays EEOE, it means that the expected algorithmic information necessary to explain the networked behaviour of a node eventually starts to grow faster than the expected algorithmic information necessary to explain the isolated behavior of the same node.
That is, as the number of nodes grows toward infinity, the algorithmic complexity of the networked behaviour of node is on average increasingly larger than the algorithmic complexity of the isolated behaviour of the node.

The proof of the occurrence of EEOE in the models studied in \cite{Abrahao2019} is achieved by employing a theoretical analysis of the trade-off between the number of communication rounds and the average density of networked nodes with the maximum algorithmic complexity.
There is an optimum balance between these two quantities where, if a large enough average density of these nodes is achieved in a sufficiently small number of communication rounds, then EEOE is triggered.

Instead of the communication protocol of plain diffusion, \citet{Abrahao2018} shows that a susceptible-infected-susceptible (SIS) contagion scheme \cite{Pastor-Satorras2001,Pastor-Satorras2002} in algorithmic networks with a power-law degree distribution is also sufficient for triggering EEOE.
In \cite{Abrahao2020}, it is shown that a slight modification in the communication protocol of plain diffusion from \cite{Abrahao2019} is sufficient for enabling the whole algorithmic network to synergistically solve problems in a higher computational class than the computational class of its individual nodes.
Not only about EEOE and synergy, the models in \cite{Abrahao2018} and \cite{Abrahao2020} also reveal that, indeed, complex networks dynamics can be employed in order to irreducibly increase the computational power of the whole network.
Thus, this approach to distributed computing formalizes what one may call ``network-inspired computing''.

As we mentioned in Section~\ref{sectionExamplesofAOIEinevolutionarysystems}, it was shown in \cite{Abrahao2015,Abrahao2016} that the evolutionary open-endedness from \cite{Chaitin2012,Chaitin2013} also applies to the resource-bounded case.
Regarding the EEOE from \cite{Abrahao2018,Abrahao2019}, further research is still needed for establishing how a resource-bounded version of the EEOE in algorithmic networks unfolds mathematically.

Unlike Theorem~\ref{thmAOIEinevolution}, since converting the result from collections of TMs into the dynamical system version is not as straightforward as in Section~\ref{sectionExamplesofAOIEinevolutionarysystems}---though the reverse is easy---we have chosen to demonstrate Theorem~\ref{thmAOIEinalgorithmicnetworks} already in the dynamical system variant of AOIE, satisfying Definition~\ref{defAOIE}.
To this end, we slightly extend our notation to encompass the case of \emph{macro-level} dynamical systems that are composed of other, \emph{micro-level}, dynamical systems.
Let $ \mathfrak{S} $ denote a FDDDS from which each (macro-level) state $ \mathfrak{S} \upharpoonright_{ x }^{ x }  $ at time instant $ x $ is a fixed arrangement of all the (micro-level) states $ \mathbf{S}_i \upharpoonright_{ x }^{ x } $ with $ 1 \leq i \leq N $, where $ x $ is an arbitrary time instant and $ N $ is the total number of dynamical systems in the form $ \mathbf{S}_i $ that composes $ \mathfrak{S} $.
Formally, we define $ \mathfrak{S} $ as an encoded isomorphic copy of the (relational) structure $ \mathscr{S} = \left( \left\{ \mathbf{S}_i \middle\vert 1 \leq i \leq N \right\} , R_1 , \dots , R_k \right) $ \cite{Hodges1993,Bresciani2019} such that, for every $ 1 \leq i \leq N $, each $ \mathbf{S}_i $ is an element of the domain of the relations $  R_1 , \dots , R_k $ and the way the micro-level FDDDSs $ \mathbf{S}_i $ is arranged is defined by an arbitrarily fixed encoding of the atomic diagram of the very relations $ R_k $ with $ 1 \leq k < \infty $.
So, each FDDDS $ \mathbf{S}_i $ is a constituent part of the larger FDDDS $ \mathfrak{S} $ and the collection of all $ \mathbf{S}_i $ that is organized/structured by the relations $  R_1 , \dots , R_k $ defines the entire $ \mathfrak{S} $.
Algorithmic information distortions due to isomorphisms as found, for example, in multidimensional networks \cite{Abrahao2021a,Abrahao2021b} do not impact the overall algorithmic information content of $ \mathfrak{S} $ because here we are fixing the encoding of the atomic diagram for any comparison between algorithmic complexities \cite{Abrahao2020a}.
For example, in case $ \mathfrak{S} $ is a bidimensional FDDDS, each state of $ \mathfrak{S} $ at time instant $ t $ can be represented as a matrix in which each entry represents a single state of $ \mathbf{S}_i $ at time instant $ t $, where $ N = m n $ and $ m $ is the number of rows and $ n $ is the number of columns.
The choice of the arrangement or structure of the micro-level FDDDSs is arbitrary and, as long as this choice is fixed for comparing $ \mathfrak{S} $ under distinct situations or conditions, it does not change our final results.

In this way, we can further explore the necessity of extra algorithmic information at a certain stage of a system when the macro-level dynamical system has reached a sufficiently large size.

The \emph{main idea} underlying Theorem~\ref{thmAOIEinalgorithmicnetworks} is that 
for a sufficiently large population/network size, the state space trajectory of $ \mathfrak{S}' $ after time instant $ t $ (which corresponds to the simulation of the population of nodes running networked) demands more algorithmic information to be computed, \emph{on average}, than the FOS is able to process after the observation of the state space trajectory of $ \mathfrak{S} $ (which corresponds to the simulation of the population of isolated nodes) until time instant $ t $.
The key steps of the following proof are to convert each node's computation into the respectively equivalent FDDDSs, as we saw in Section~\ref{sectionFOS}. 
First, we construct the dynamical system $ \mathbf{S}_i $ until time instant $ t $, corresponding to the node $ o_i $ running isolated from each other.
Secondly, we construct the dynamical system $ \mathbf{S}'_i $ from time instant $ t + 1 $ until $ t' $, corresponding to the node $ o_i $ running networked.
Thus, a node $ o_i $ receiving information from its neighbour nodes (according to the network topology) is equivalent to the FDDDS $ \mathbf{S}'_i $ being algorithmically perturbed by its neighbour FDDDSs (also according to the same network topology).
Then, we combine these FDDDSs $ \mathbf{S}_i $ and $ \mathbf{S}'_i $ in order to form the macro-level dynamical systems $ \mathfrak{S} $ and $ \mathfrak{S}' $, which refer to the isolated and networked cases, respectively.
In other words, $ \mathfrak{S}' $ is composed of a population of randomly generated FDDDSs that can perturb each other according to the network topology.
On the other hand, although $ \mathfrak{S} $ is composed of the same population as $ \mathfrak{S}' $ does, no FDDDS in $ \mathfrak{S} $ can perturb other FDDDSs in $ \mathfrak{S} $.

\begin{theorem}\label{thmAOIEinalgorithmicnetworks}
	Let $ \mathfrak{S} $ be the macro-level FDDDS whose decision problem is Turing equivalent to calculating the final output of every \emph{isolated} node $ o_i $, which belongs to a population of $ N $ isolated nodes as in \cite{Abrahao2019}, where $ 1 \leq i \leq N $.
	Let $ \mathfrak{S}' $ be the macro-level FDDDS whose decision problem is Turing equivalent to calculating the final output of every \emph{networked} node $ o_i $, which belongs to an algorithmic network $ \mathfrak{N} $ that displays EEOE as in \cite{Abrahao2019}, and the set of nodes of $ \mathfrak{N} $ is the same population in the former isolated case. 
	Then, 
	for every $ \mathcal{O} $, for every time instant $ t $, and for every $ c_\mathbf{I} , c_{ \mathcal{O} } , c_e > 0 $, there is $ t_e \in T $ such that, for every $ t' \geq t_e $, one has it that 
	\begin{equation*}
		\mathbf{E}_{  i \leq N }\left( \mathbf{ I_{ ac } }\left( \left< \mathbf{S}_i \upharpoonright_{ 0 }^{ t } ,  \mathbf{S}'_i \upharpoonright_{ t + 1 }^{ t' } \right> \, \middle\vert \, \left< w , \left( \mathcal{O} \upharpoonright_{ 0 }^{ t } ,  \mathcal{O}' \upharpoonright_{ t + 1 }^{ t + m } \right) \right> \right) \right)
		 > 
		 c_\mathbf{I} + c_{ \mathcal{O} } + c_e
	\end{equation*}
	holds with probability arbitrarily close to $1$ as $ N \to \infty $,
	where:
	$ t' \geq t + m $, $ m \geq 1 $, $ w $ is the bit string in the first tape of $ O $ that satisfies the Definition~\ref{defObservationprinciple} at time instant $ t $;
	$ \mathbf{S}_i $ and $ \mathbf{S}'_i $ are the micro-level FDDDSs that simulate the node $ o_i $ in the isolated and the networked cases, respectively;
	and the collection of all $ \mathbf{S}_i $ and $ \mathbf{S}'_i $, for every $ i > 0 $, form the macro-level FDDDSs $ \mathfrak{S} $ and $ \mathfrak{S}' $, respectively.
	
	\begin{proof}
		Let $ \mathfrak{N} $ be an algorithmic network studied in \cite{Abrahao2019} that displays EEOE.
		Since every node $ o_i $ is a randomly generated program (or TM), we need to first construct one equivalent dynamical system that simulates the \emph{networked} behaviour of the node $ o_i $ when the dynamical system is interacting with (i.e., perturbing and being perturbed by) its neighbours, and construct another one that simulates the isolated behaviour of the node $ o_i $ when the dynamical system is \emph{not} interacting with any of its neighbours.
		The isolated case is easily obtained by just replacing each node $ o_i $ with a FDDDS $ \mathbf{S}_i $ that has the same e.r. as $ A $ such that the initial state $ \left( {\mathbf{S}_i} \upharpoonright_{ 0 }^{ 0 } \right) $ corresponds to the initial configuration of $ A $ simulating
		$ M\left( M'\left( P_{iso}\left( o_i , c \right) , \emptyset \right) \right) $, 
		where 
		the universal FDDDS $ A $ and TMs $ M $ and $ M' $ are exactly those defined in the proof of Lemma~\ref{lemmaSequence computabilityofO}.
		Therefore, there is $ h \geq 0 $ such that, for every isolated $ o_i $, we have it that 
		\begin{equation}\label{equationIsolatedcase}
			\mathbf{K}\left(  \mathbf{S}_i \upharpoonright_{ 0 }^{ t }  \right)
			=
			\mathbf{K}\left( \mathbf{U}\left( P_{iso}\left( o_i , c \right) \right) \right)
			\pm \mathbf{O}(1)
			\text{ ,}
		\end{equation}
		where $ t = c + h - 1 $.
		In order to obtain the networked case, we then need to demonstrate that 
		there is an equivalent AP $ \mathcal{P}_{ \left( o_i , c' \right) } $ that transforms the FDDDS $ {\mathcal{S}'_i}^{ ( c' ) } $ (which computes the output that is sent to $ o_i $'s neighbours in communication round $ c' $) into the FDDDS $ {\mathcal{S}'_i}^{ ( c' + 1 ) } $ (which computes the output that is sent to $ o_i $'s neighbours in communication round $ c' + 1 $).
		To this end, by employing the universal FDDDS $ A $ and TMs $ M $ and $ M' $ from the proof of Lemma~\ref{lemmaSequence computabilityofO}, one constructs a sequence $ \mathbf{S}'_i = \left( {\mathcal{S}'_i}^{ ( 1 ) } , \dots , {\mathcal{S}'_i}^{ ( c' ) } , \dots , {\mathcal{S}'_i}^{ ( c ) } \right) $ of FDDDSs and the program $ \mathcal{P}_{ \left( o_i , c' \right) } $ that returns the state $ \left( {\mathcal{S}'_i}^{ ( c' + 1 ) } \upharpoonright_{ t_{ (c' + 1) } }^{ t_{ (c' + 1) } } \right) $ given the state $ \left( {\mathcal{S}'_i}^{ ( c' ) } \upharpoonright_{ t_{ (c' + 1) } - 1 }^{ t_{ (c' + 1) } - 1 } \right) $ as input, 
		where: 
		the state space trajectory
		$ \left( \mathcal{S}'^{ ( 1 ) } \upharpoonright_{ t + 1 }^{ t_{ ( 2 ) } - 1 } \right) $ computes $ \mathbf{U}\left( P_{iso}\left( o_i , 1 \right) \right) $;
		the state space trajectory $ \left( {\mathcal{S}'_i}^{ ( c' ) } \upharpoonright_{ t_{ (c') } }^{ t_{ (c' + 1) } - 1 } \right) $ computes the output that is sent to $ o_i $'s neighbours in communication round $ c' $;
		the state space trajectory $ \left( {\mathcal{S}'_i}^{ ( c' + 1 ) } \upharpoonright_{ t_{ (c' + 1) } }^{ t_{ (c' + 2) } - 1 } \right) $ computes the output that is sent to $ o_i $'s neighbours in communication round $ c' + 1 $; 
		and so on.
		Note that $ c' $ is an arbitrary communication round with $ 1 \leq c' \leq c $.
		Therefore, for every networked $ o_i $,
		\begin{equation}\label{equationNetworkedcase}
			\mathbf{K}\left(  \left< {\mathcal{S}'}_i^{ ( 1 ) } \upharpoonright_{ t + 1 }^{ t_{ ( 2 ) } - 1 } , \dots , {\mathcal{S}'}_i^{ ( c' ) } \upharpoonright_{ t_{ ( c' ) } }^{ t_{ ( c' + 1 ) } - 1 } , \dots , {\mathcal{S}'}_i^{ ( c ) } \upharpoonright_{ t_{ ( c ) } }^{ t' } \right>  \right)
			=
			\mathbf{K}\left( \mathbf{U}\left( P_{net}\left( o_i , c \right) \right) \right)
			\pm \mathbf{O}(1)
			\text{ .}
		\end{equation}
		Let $ \mathfrak{S} $ and $ \mathfrak{S}' $ be the FDDDSs composed of a fixed arrangement of all the (micro-level) states of $ \mathbf{S}_i $ and $ \mathbf{S}'_i $, respectively, where $ 1 \leq i \leq N $ and $ N $ is the total number of nodes.
		Let $ \mathbf{E}_{ i \leq N }\left( \cdot \right) $ denote the average over all the constituent systems $ \mathbf{S}_i $ or $ \mathbf{S}'_i $ that composes $ \mathfrak{S} $ or $ \mathfrak{S}' $, respectively.
		We know from \cite{Abrahao2019} that Equation~\ref{equationEEOE} holds with probability arbitrarily close to $ 1 $ as $ N \to \infty $.
		Therefore, from basic inequalities in AIT and from the Definition~\ref{defObservationprinciple}, one finally achieves the proof of Theorem~\ref{thmAOIEinalgorithmicnetworks} by combining Equation~\ref{equationEEOE} with Equations~\ref{equationIsolatedcase} and \ref{equationNetworkedcase}.
	\end{proof}
\end{theorem}

Extending the result of Theorem~\ref{thmAOIEinalgorithmicnetworks} to the algorithmic networks in \cite{Abrahao2018} is straightforward, if one sets out from the method employed in the proof of Theorem~\ref{thmAOIEinalgorithmicnetworks}.
Thus we leave it up to the reader.

While the temporal (or diachronic) variant of AOIE presented in the previous Section~\ref{sectionExamplesofAOIEinevolutionarysystems} occurs over time due to the successive perturbations of the system $ \mathbf{P} $ (or $ \mathbf{S} $) by the environment,
the variant of AOIE presented in Theorem~\ref{thmAOIEinalgorithmicnetworks} occurs due to the interaction (in the form of perturbations) between the micro-level systems $ \mathbf{S}'_i $ as the number of these micro-level systems contained in the macro-level system $ \mathfrak{S}' $ increases.
Although there may be a FOS that can compute the expected behaviour of an isolated micro-level system, the AOIE guarantees that this will eventually cease to happen as the size of the macro-level system becomes sufficiently large.
The existence of the time instant $ t_e $ in Theorem~\ref{thmAOIEinalgorithmicnetworks} ensures that, even if the FOS can computably predict the expected behaviour of an isolated micro-level system that belongs to $ \mathfrak{S} $, there is a phase transition for which, if the expected behaviour of a micro-level system of $ \mathfrak{S}' $ is not emergent vis-\`{a}-vis a certain FOS, then it will start to be emergent once the number of micro-level systems in $ \mathfrak{S}' $ is sufficiently large.
Thus, the sort of process that gives rise to the AOIE in Theorem~\ref{thmAOIEinalgorithmicnetworks} differs from the one in Theorem~\ref{thmAOIEinevolution} in the same manner as the holistic variant of emergence differs from the temporal (or diachronic) one.
For this reason, AOIE in Theorem~\ref{thmAOIEinalgorithmicnetworks} falls under the \emph{holistic} variant of emergence.
Hence, one can adopt the convention of referring to the \emph{emergent open-endedness} proved in Theorem~\ref{thmAOIEinalgorithmicnetworks} as \emph{asymptotically observer-independent holistic open-endedness}.

\emph{Downward (or top-down) causation}  is usually described in the literature as a type of process in which the global (or macro-level) dynamics of the system as a ``whole'' gains causal efficacy over the micro-level systems (or parts)
\cite{O'Connor2020,Davies2008,Chalmers2008}.
The holistic variant of AOIE in Theorem~\ref{thmAOIEinalgorithmicnetworks} demonstrates that for sufficiently large $ \mathfrak{S}' $, the expected behaviour of a networked micro-level system $ \mathbf{ S }'_i $ is overruled by the algorithmic-informational dynamics of the algorithmic perturbations produced according to the network topology.
This occurs because the algorithmic information of the dynamics of an isolated $ \mathbf{ S }_i $ will eventually be insufficient for computing the networked behaviour of $ \mathbf{ S }'_i $, while the total algorithmic information shared through the network is.
In this regard, Theorem~\ref{thmAOIEinalgorithmicnetworks} gives a proof of the existence of expected downward causation in FDDDSs (or in networked computable systems).
It also offers the advantage of this expected downward causation being independent of the observer's formal knowledge at the asymptotic limit. 


\subsubsection{Weak, intermediate, or strong emergence}\label{sectionWeakorStrongemergence}

In a broad sense, if weak emergence is characterised by phenomena that are \emph{in principle} deducible or derivable from simple initial or micro-level conditions, but that appear unexpected at a higher coarse-grained level due to a lack of information, resources, or knowledge, one can classify the ODE in Section~\ref{sectionWeakemergentalgorithmicinformation} as weak emergence.
This agrees with the approach to weak emergence as the unexpectedly complex behavior in \cite{Chalmers2008}, as explanatory incompressibility in \cite{Bedau2010}, and as the type 0 and 1 weak emergence in \cite{Bar-Yam2004}.
Indeed, since there is always the possibility of another existing observer to which the phenomenon ceases to appear emergent, then the emergence in Definition~\ref{defObserverdependentemergence} is, ``\emph{in principle}'', deducible or derivable at the same time that there are observers for which the emergent behaviour is ``truly'' incompressible and relatively uncomputable.
The term ``truly'' is employed here in the precise sense that such an incompressibility or relative uncomputability does not depend on the method chosen to measure the information content, on the errors or distortions in the act of observing, or on the algorithmic-informational cost to process the observed system in accordance with the observer's formal knowledge.

On the other hand, classifying the AOIE studied in Section~\ref{sectionAOIE} is not so easy. 
The crux of the matter lies not quite in the notion of reducibility, derivability, or predictability (as in our case they have a formal unambiguous translation into sufficient algorithmic information), but in the term ``in principle''.
If ``in principle'' means that the phenomenon should remain emergent for every formal observer system that belongs to the same computational class as the observed systems, then AOIE could be interpreted as a type of strong emergence.
This is because for every formal observer system of the same computational class (i.e., in the same Turing degree or in the same running time complexity class) as the observed systems, the behaviour of an observed system that satisfies Definition~\ref{defAOIE} will eventually cease to be computable or predictable in the long run.

In this sense, since AOIE implies ODE for infinitely many time steps in the future, the Church-Turing hypothesis entails that a system displaying AOIE (and, in this case, strong emergence) will always be understood as displaying ODE (and therefore the above weak emergence), while in fact never ceasing to display ODE (or weak emergence) for any possible observer.
In other words, under the Church-Turing hypothesis, if AOIE is considered strong emergence, then this type of strongly emergent phenomenon is a pseudoparadoxical type of emergent phenomenon that is always mathematically understood as weak emergence by us, while in fact displaying strong emergence (if an hypothetical observer could know the point of view of every observer).

Another form of strong emergence has been described as the ultimate necessity of novel fundamental powers or laws to scientifically explain the macro-level behaviour of a system \cite{O'Connor2020}.
In the context of FDDDSs or computable systems, AOIE offers a proof of this necessity, but now formally expressed as the never-ending necessity of new axioms (or new algorithmic information).
Due to the presence of expected downward causation in the AOIE of Theorem~\ref{thmAOIEinalgorithmicnetworks}, one can also successfully argue that the systems $ \mathfrak{S}' $ satisfy the type 2 strong emergence in \cite{Bar-Yam2004}.

However, if ``in principle'' does not constrain the computational class of the observer, then AOIE can be brought back to the weak case.
This is because although no (finite) formal axiomatic theory held by the observer can compute the observed system in the long run, there may be \emph{oracle} observers that can, if the e.r. of the observer itself belongs to higher Turing degrees.
For example, it is true that both the sequence $ \mathbf{P} $ of TMs in Theorem~\ref{thmAOIEinevolution} and the macro-level FDDDS $ \mathfrak{S}' $ in Theorem~\ref{thmAOIEinalgorithmicnetworks} cannot be computed by formal observer systems at the asymptotic limit, but both can be computed by an oracle machine of Turing degree $ \mathbf{ 0' } $.
In other words, if one allows observers to have access to an infinite source of algorithmic information, e.g., by filling out the \emph{infinite} second tape of $ \mathcal{O} $ with a halting probability (or Chaitin's Omega number) \cite{Chaitin1975,Calude2002},  
there are systems that satisfy Definition~\ref{defAOIE} at the same time that they are relatively computable by a \emph{special} observer.
Thus, in cases where one believes in the existence of strong emergence that resists ontological characterisation in terms of physical or informational causal efficacies, such as the emergence of qualia in the conscious mind \cite{Chalmers2008,O'Connor2020}, it becomes consistent to classify AOIE as a type of weak emergence.

Although not constraining the computational class of the observer may seen reasonable, one is inherently assuming that there are ``special'' observers that belong to a higher computational class than that of all the other systems that can be observed by them, an assumption which \textit{per se}  is just another type of constraint to be applied to Definition~\ref{defAOIE}.
One way to avoid this assumption, while still being consistent with the fact that AOIE from Definition~\ref{defAOIE} is a stronger form of emergence than what is usually considered to be weak emergence (which generally falls under the ODE from Definition~\ref{defObserverdependentemergence}) is to classify AOIE as a type of \emph{intermediate emergence} \cite{Chalmers2008}.
This kind of terminology has been proposed by \citet{Chalmers2008} to deal with a type of emergence that arises from a fundamental epistemological limitation, given the known physical laws at the time of observation.
In this sense, intermediate emergence is predicated upon the existence of the unbridgeable incompleteness of the observer's knowledge, so that even ``in principle'' one would not be able to deduce the macro-level complex behaviour, which is still ``in principle'' determined by irreducible new laws that one always needs to devise or discover in the future.
Regarding incompleteness, emergence as a consequence of uncomputability was also proposed by \citet{Cooper2009}.

Classifying AOIE as intermediate emergence implies an underlying assumption of the existence (or scientific pertinence) of a stronger form of emergence, which is an open problem. Nevertheless, we consider both hypotheses (i.e., with or without ``special'' observers) of mathematical and scientific relevance and worth pursuing.
For present purposes, since we have only dealt with formal axiomatic theories and not with physical, chemical, or biological theories in general, we hold on to what our theoretical results imply, and therefore we avoid the claim of classifying AOIE as either weak, strong, or intermediate emergence.
What we have shown is that, restricted to the context of algorithmic information dynamics, FDDDSs, computable systems, and FOSs, the AOIE is the \emph{strongest} form of emergence that formal axiomatic theories can attain.
Algorithmic information and algorithmic randomness have demonstrated and captured fundamental properties that underlie the incompleteness of formal theories and the limits of mathematics \cite{Chaitin1987,Calude2002,Downey2010}. 
Thus, within the scope of this article, it may not come as a surprise that algorithmic information theory was the key to formalising emergence up to the limits that our formal mathematical knowledge can grasp.

\section{Conclusion}\label{sectionConclusion}

Within the scope of algorithmic information dynamics, this article studies the fundamental role that algorithmic information plays in the act of observing and in the occurrence of emergent phenomena in discrete deterministic dynamical systems and computable systems.

We have formalised the act of observing a system as mutual perturbations occurring between the observer (which is itself a system) and the observed system.
Formal observer systems are systems that previously know a formal axiomatic theory, which they can apply in order to compute the future behaviour of the observed system. 
As a consequence, we demonstrate that a (finite discrete deterministic dynamical) system displaying emergent behaviour with respect to an observer constitutes a type of emergence of algorithmic information that is invariant and minimal.
Although it depends on the observer, this emergence is robust in the face of variations of the arbitrarily chosen method of measuring irreducible information content, errors (or distortions) in the very act of observing, and variations of the algorithmic-informational cost of processing the information gathered from the observed system in accordance with the observer's formal knowledge.
Thus, this type of emergence is called \emph{observer-dependent emergence} (ODE).

Then, we investigated the unbounded increase of emergent algorithmic information, which defines a type of emergence that we call \emph{asymptotically observer-independent emergence} (AOIE).
In addition to the above invariance, minimality, and robustness, any formal axiomatic theory that any formal observer system might devise will eventually fail to compute or predict the behaviour of a system that displays AOIE.
Thus, although each formal observer system retains its own subjectivity as in the above ODE, AOIE defines a type of emergence that outdoes any subjectivity at the asymptotic limit.

We demonstrated that there is an abstract evolutionary model that displays the temporal (or diachronic) variant of AOIE, which guarantees that no formal observer system is able to always compute the behaviour of evolutionary computable systems in the long run.
We also demonstrated that there is an abstract model for networked systems that displays the holistic variant of AOIE, which guarantees that no formal observer system is able to always compute the expected behaviour of a micro-level subsystem as the size of the macro-level system becomes sufficiently large.

We also compared the ODE and AOIE studied in this article with weak and strong emergence in the literature.
Depending on the interpretation of the term ``in principle'' in the usual definitions of weak and strong emergence, AOIE can be classified as weak, intermediate, or strong emergence.
In any event, the results of the present article show that, within the context of finite discrete deterministic dynamical systems, or computable systems, AOIE is the strongest version of emergence that formal axiomatic theories can grasp or capture.
Whether this claim can be extended to other physical systems and physical theories is a problem that needs further discussion and future research.
Nevertheless, given the relevance of formal axiomatic theories in mathematics and science in general, we consider the strength of AOIE demonstrated in this article to be remarkable.

%
%
%
%
%
%
%




\begin{thebibliography}{9}
	\bibitem[Abrahão(2015)]{Abrahao2015} Abrahão FS. 2015 \textit{Metabiologia, Subcomputação e Hipercomputação: em direção a uma teoria geral de evolução de sistemas}. Ph.D. thesis, Federal University of Rio de Janeiro (UFRJ), Brazil, Rio de Janeiro. Language: Portuguese. Available at \href{http://objdig.ufrj.br/10/teses/832593.pdf}{http://objdig.ufrj.br/10/teses/832593.pdf}
	
	
	\bibitem[Abrahão(2016)]{Abrahao2016} Abrahão FS. 2016 The ‘paradox’ of computability and a recursive relative version of the Busy Beaver function. In \textit{Information and Complexity} (eds C Calude, M Burgin), pp. 3–15. Singapure: World Scientific Publishing. \href{http://doi.org/10.1142/9789813109032_0001}{(doi:10.1142/9789813109032{\_}0001)}
	
	\bibitem{Abrahao2016b} Abrahão FS. 2016 Emergent algorithmic creativity on networked Turing machines. In \textit{The 8th International Workshop on Guided Self-Organization at the Fifteenth International Conference on the Synthesis and Simulation of Living Systems} (ALIFE), Cancún. 
	
	\bibitem{Abrahao2020} Abrahão FS, D’Ottaviano ÍML, Wehmuth K, Dória FA, Ziviani A. 2020 Learning the undecidable from networked systems. In \textit{Unravelling Complexity} (eds S Wuppuluri, FA Doria), World Scientific Publishing. \href{https://doi.org/10.1142/9789811200076_0009}{(doi:10.1142/9789811200076{\_}0009)}
	
	\bibitem{Abrahao2020a} Abrahão FS, Wehmuth K, Zenil H, Ziviani A. 2020 Algorithmic information and incompressibility of families of multidimensional networks. arXiv Preprints.
	Available at \href{http://arxiv.org/abs/1810.11719}{http://arxiv.org/abs/1810.11719}
	
	\bibitem{Abrahao2021a} Abrahão FS, Wehmuth K, Zenil H, Ziviani A. 2021 An Algorithmic Information Distortion in Multidimensional Networks. In \textit{Complex Networks and Their Applications IX} (eds RM Benito, C Cherifi, H Cherifi, E Moro, LM Rocha, M Sales-Pardo), pp. 520–531. Cham: Springer International Publishing. \href{http://doi.org/10.1007/978-3-030-65351-4_42}{(doi:10.1007/978-3-030-65351-4{\_}42)}
	
	\bibitem{Abrahao2021b} Abrahão FS, Wehmuth K, Zenil H, Ziviani A. 2021 Algorithmic Information Distortions in Node-Aligned and Node-Unaligned Multidimensional Networks. \textit{Entropy} \textbf{23}, 835. (doi:10.3390/e23070835)
	\href{http://doi.org/10.3390/e23070835}{(doi:10.3390/e23070835)}
	
	
	\bibitem[Abrahão et al(2018)]{Abrahao2018} Abrahão FS, Wehmuth K, Ziviani A. 2018 Emergent Open-Endedness from Contagion of the Fittest. \textit{Complex Systems} \textbf{27}. \href{http://doi/org/10.25088/ComplexSystems.27.4.369}{(doi:10.25088/ComplexSystems.27.4.369)}
	
	\bibitem{Abrahao2019} Abrahão FS, Wehmuth K, Ziviani A. 2019 Algorithmic networks: Central time to trigger expected emergent open-endedness. \textit{Theoretical Computer Science} \textbf{785}, 83–116. \href{http://doi.org/10.1016/j.tcs.2019.03.008}{(doi:10.1016/j.tcs.2019.03.008)}
	
	
	\bibitem{Adams2017a} Adams AM, Berner A, Davies PCW, Walker SI. 2017 Physical Universality, State-Dependent Dynamical Laws and Open-Ended Novelty. \textit{Entropy} \textbf{19}, 461. \href{http://doi.org/10.3390/e19090461}{(doi:10.3390/e19090461)}
	
	
	\bibitem{Adams2017} Adams A, Zenil H, Davies PCW, Walker SI. 2017 Formal Definitions of Unbounded Evolution and Innovation Reveal Universal Mechanisms for Open-Ended Evolution in Dynamical Systems. \textit{Scientific Reports} \textbf{7}, 997. \href{http://doi.org/10.1038/s41598-017-00810-8}{(doi:10.1038/s41598-017-00810-8)}
	
	\bibitem{Barabasi2009} Barabási A-L. 2009 Scale-Free Networks: A Decade and Beyond. \textit{Science} \textbf{325}, 412–413. \href{https://doi.org/10.1126/science.1173299}{(doi:10.1126/science.1173299)}
	
	
	\bibitem{Barabasi2016} Barabási A-L. 2016 \textit{Network Science}. 1st edn. USA: Cambridge University Press. 
	
	
	
	\bibitem{Bar-Yam2004} Bar-Yam Y. 2004 A mathematical theory of strong emergence using multiscale variety. \textit{Complexity} \textbf{9}, 15–24. \href{hhtp://doi.org/10.1002/cplx.20029}{(doi:10.1002/cplx.20029)}
	
	\bibitem{Becher2002} Becher V, Figueira S. 2002 An example of a computable absolutely normal number. \textit{Theoretical Computer Science} \textbf{270}, 947–958. \href{http://doi.org/10.1016/S0304-3975(01)00170-0}{(doi:10.1016/S0304-3975(01)00170-0)}
	
	
	
	\bibitem[Bedau(1997)]{Bedau1997} Bedau MA. 1997 Weak Emergence. \textit{Philosophical Perspectives} \textbf{11}, 375–399. \href{hhtp://doi.org/10.1111/0029-4624.31.s11.17}{(doi:10.1111/0029-4624.31.s11.17)}
	
	\bibitem[Bedau(2010)]{Bedau2010} Bedau MA. 2010 Weak Emergence and Context-Sensitive Reduction. In \textit{Emergence in Science and Philosophy} (eds A Corradini, T O’Connor), Routledge. \href{https://doi.org/10.4324/9780203849408}{(doi:10.4324/9780203849408)}
	
	\bibitem{Bresciani2019} Bresciani E, D’Ottaviano ÍML. 2019 Basic concepts of systemics. In \textit{Systems, Self-Organisation and Information: an interdisciplinary perspective} (eds PJ Alfredo, WA Pickering, RR Gudwin), Abingdon, Oxon; New York, NY: Routledge, 2019.: Routledge.
	
	
	\bibitem{Burgin2009} Burgin M. 2009 Theory of Information: Fundamentality, Diversity and Unification. World Scientific. \href{http://doi.org/10.1142/7048}{(doi:10.1142/7048)}
	
	
	
	\bibitem{Calude2002} Calude CS. 2002 \textit{Information and Randomness: An algorithmmic perspective}. 2nd edn. Springer-Verlag. Available at \href{https://www.springer.com/br/book/9783540434665}{https://www.springer.com/br/book/9783540434665}
	
	\bibitem[Chaitin(1975)]{Chaitin1975} Chaitin GJ. 1975 A Theory of Program Size Formally Identical to Information Theory. \textit{Journal of the ACM} \textbf{22}, 329–340. \href{https://doi.org/10.1145/321892.321894}{(doi:10.1145/321892.321894)}
	
	
	\bibitem{Chaitin1987} Chaitin G. 2004 \textit{Algorithmic Information Theory}. 3rd edn. Cambridge University Press. \href{http://doi.org/10.1017/CBO9780511608858}{(doi:10.1017/CBO9780511608858)}
	
	\bibitem[Chaitin(2012)]{Chaitin2012} Chaitin G. 2012 Life as evolving software. In \textit{A Computable Universe: Understanding and Exploring Nature as Computation} (ed H Zenil), pp. 1–23. Singapure: World Scientific Publishing. \href{http://doi.org/10.1142/9789814374309_0015}{(doi:10.1142/9789814374309{\_}0015)}
	
	\bibitem{Chaitin2013} Chaitin GJ. 2012 \textit{Proving Darwin: making biology mathematical}. New York: Pantheon Books. 
	
	
	
	\bibitem{Chaitin2018} Chaitin VMFG, Chaitin GJ. 2018 A Philosophical Perspective on a Metatheory of Biological Evolution. In \textit{The Map and the Territory: Exploring the Foundations of Science, Thought and Reality} (eds S Wuppuluri, FA Doria), pp. 513–532. Cham: Springer International Publishing. \href{https://doi.org/10.1007/978-3-319-72478-2_29}{(doi:10.1007/978-3-319-72478-2{\_}29)}
	
	
	\bibitem[Chalmers(2008)]{Chalmers2008} Chalmers DJ. 2008 Strong and Weak Emergence. In \emph{The Re-Emergence of Emergence} (eds P Clayton, PCW Davies), Oxford: Oxford University Press.
	 \href{hhtp://doi.org/10.1093/acprof:oso/9780199544318.003.0011}{(doi:10.1093/acprof:oso/9780199544318.003.0011)}
	
	
	
	\bibitem[Cooper(2009)]{Cooper2009} Cooper SB. 2009 Emergence as a computability-theoretic phenomenon. \textit{Applied Mathematics and Computation} \textbf{215}, 1351–1360. \href{http://doi.org/10.1016/j.amc.2009.04.050}{(doi:10.1016/j.amc.2009.04.050)}
	
	\bibitem{Cover2005} Cover TM, Thomas JA. 2005 \textit{Elements of Information Theory}. Hoboken, NJ, USA: John Wiley \& Sons, Inc.
	\href{http://doi.org/10.1002/047174882X}{(doi:10.1002/047174882X)}
	
	\bibitem{Davies2008} Davies PCW. 2008 The Physics of Downward Causation. In The \textit{Re-Emergence of Emergence} (eds P Clayton, PCW Davies), Oxford: Oxford University Press. \href{http://doi.org/10.1093/acprof:oso/9780199544318.003.0002}{(doi:10.1093/acprof:oso/9780199544318.003.0002)}
	
	\bibitem{Diestel2017} Diestel R. 2017 \textit{Graph Theory}. 5th edn. Springer-Verlag.
	
	\bibitem{Downey2010} Downey RG, Hirschfeldt DR. 2010 Algorithmic Randomness and Complexity. New York, NY: Springer New York. \href{http:/doi.org/10.1007/978-0-387-68441-3}{(doi:10.1007/978-0-387-68441-3)}
	
	\bibitem{Fernandez2014} Fernández N, Maldonado C, Gershenson C. 2014 Information Measures of Complexity, Emergence, Self-organization, Homeostasis, and Autopoiesis. In \textit{Guided Self-Organization: Inception} (ed M Prokopenko), pp. 19–51. Springer, Berlin, Heidelberg. (\href{http://doi.org/10.1007/978-3-642-53734-9_2}{doi:10.1007/978-3-642-53734-9{\_}2})
	
	
	
	
	\bibitem[Hernandez-Orozco et al(2018)]{Hernandez-Orozco2018} Hernández-Orozco S, Hernández-Quiroz F, Zenil H. 2018 Undecidability and Irreducibility Conditions for Open-Ended Evolution and Emergence. \textit{Artificial Life} \textbf{24}, 56–70. (\href{http://doi.org/10.1162/ARTL_a_00254}{doi:10.1162/ARTL{\_}a{\_}00254})
	
	\bibitem{Hernandez-Orozco2018a} Hernández-Orozco S, Kiani NA, Zenil H. 2018 Algorithmically probable mutations reproduce aspects of evolution, such as convergence rate, genetic memory and modularity. \textit{Royal Society Open Science} \textbf{5}, 180399. (\href{http://doi.org/10.1098/rsos.180399}{doi:10.1098/rsos.180399})
	
	
	\bibitem{Hodges1993} Hodges W. 1993 Model Theory. 1st edn. Cambridge University Press.
	(\href{http://doi.org/10.1017/CBO9780511551574}{doi:10.1017/CBO9780511551574})
	
	
	\bibitem{Lewis1997} Lewis H, Papadimitriou CH. 1997 \textit{Elements of the Theory of Computation}. 2nd edn. Prentice-Hall. 
	
	\bibitem{Lewis2009} Lewis TG. 2009 Network Science. Hoboken, NJ, USA: John Wiley \& Sons, Inc.
	(\href{http:/doi.org/10.1002/9780470400791}{doi:10.1002/9780470400791})
	
	\bibitem{Li1997} Li M, Vitányi P. 2019 \textit{An Introduction to Kolmogorov Complexity and Its Applications}. 4th edn. Cham: Springer, Cham. (\href{http:/doi.org/10.1007/978-3-030-11298-1}{doi:10.1007/978-3-030-11298-1})
	
	
	\bibitem{Lizier2018} Lizier J, Bertschinger N, Jost J, Wibral M. 2018 Information Decomposition of Target Effects from Multi-Source Interactions: Perspectives on Previous, Current and Future Work. \textit{Entropy} \textbf{20}, 307. (\href{http://doi.org/10.3390/e20040307}{doi:10.3390/e20040307})
	
	\bibitem{Mediano2019} Mediano PAM, Rosas F, Carhart-Harris RL, Seth AK, Barrett AB. 2019 \textit{Beyond integrated information: A taxonomy of information dynamics phenomena}. arXiv:1909.02297
	
	\bibitem{Michail2018} Michail O, Spirakis PG. 2018 Elements of the theory of dynamic networks. \textit{Communications of the ACM} \textbf{61}, 72–72. (\href{https://doi.org/10.1145/3156693}{doi:10.1145/3156693})
	
	
	\bibitem{Mitchell2009} Mitchell M. 2009 \textit{Complexity: A Guided Tour}. Oxford University Press. Available at \href{https://www.google.com.br/books/edition/Complexity/bbN-6aDFrrAC?hl=en&gbpv=0&kptab=overview}{ISBN 9780195124415}.
	
	\bibitem{Neary2006} Neary T, Woods D. 2006 P-completeness of Cellular Automaton Rule 110. In \textit{Automata, Languages and Programming} (eds M Bugliesi, B Preneel, V Sassone, I Wegener), pp. 132–143. Berlin, Heidelberg: Springer Berlin Heidelberg. (\href{http://doi.org/10.1007/11786986_13}{doi:10.1007/11786986{\_}13})
	
	
	\bibitem{Nicolis2009} Nicolis G, Nicolis C. 2009 \textit{Foundations of complex systems}. World Scientific Publishing Co. Pte. Ltd. (\href{http://doi.org/10.1017/S1062798709000738}{doi:10.1017/S1062798709000738})
	
	\bibitem[O'Connor(2020)]{O'Connor2020} O'Connor T. 2020 Emergent Properties. In \textit{The Stanford Encyclopedia of Philosophy} (ed EN Zalta), Metaphysics Research Lab, Stanford University. Available at \href{https://plato.stanford.edu/archives/fall2020/entries/properties-emergent/}{https://plato.stanford.edu/archives/fall2020/entries/properties-emergent/}
	
	
	\bibitem{Pastor-Satorras2001} Pastor-Satorras R, Vespignani A. 2001 Epidemic spreading in scale-free networks. \textit{Physical Review Letters} \textbf{86}, 3200–3203. (\href{https://doi.org/10.1103/PhysRevLett.86.3200}{doi:10.1103/PhysRevLett.86.3200})
	
	\bibitem{Pastor-Satorras2002} Pastor-Satorras R, Vespignani A. 2002 Immunization of complex networks. \textit{Physical Review E - Statistical, Nonlinear, and Soft Matter Physics} \textbf{65}, 036104. (\href{https:/doi.org/10.1103/PhysRevE.65.036104}{doi:10.1103/PhysRevE.65.036104})
	
	\bibitem[Polani(2003)]{Polani2003} Polani D. 2003 Measuring Self-Organization via Observers. In \textit{Advances in Artificial Life} (eds W Banzhaf, J Ziegler, T Christaller, P Dittrich, JT Kim), pp. 667–675. Berlin, Heidelberg: Springer Berlin Heidelberg.
	(\href{https:/doi.org/10.1007/978-3-540-39432-7_72}{doi:10.1007/978-3-540-39432-7{\_}72})
	
	\bibitem{Prokopenko2009} Prokopenko M, Boschetti F, Ryan AJ. 2009 An information-theoretic primer on complexity, self-organization, and emergence. \textit{Complexity} \textbf{15}, 11–28. (\href{http://doi.org/10.1002/cplx.20249}{doi:10.1002/cplx.20249})
	
	\bibitem{Rich2007} Rich EA. 2007 \textit{Automata Theory and Applications}. USA: Prentice-Hall, Inc. 
	
	
	
	\bibitem{Rosas2020} Rosas FE, Mediano PAM, Jensen HJ, Seth AK, Barrett AB, Carhart-Harris RL, Bor D. 2020 Reconciling emergences: An information-theoretic approach to identify causal emergence in multivariate data. \textit{PLoS Comput Biol} \textbf{16}, e1008289.
	(\href{http://doi.org/10.1371/journal.pcbi.1008289}{doi:10.1371/journal.pcbi.1008289})
	
	\bibitem{Shalizi2001} Shalizi CR. 2001 \textit{Causal architecture, complexity and self-organization in time series and cellular automata}. PhD Thesis, the University of Wisconsin - Madison.
	
	\bibitem[Walker and Davies(2012)]{Walker2012} Walker SI, Davies PCW. 2012 The Algorithmic Origins of Life. \textit{Journal of The Royal Society Interface} \textbf{10}, 20120869–20120869. (\href{http://doi.org/10.1098/rsif.2012.0869}{doi:10.1098/rsif.2012.0869})
	
	\bibitem{Wolfram2002} Wolfram S. 2002 \textit{A new kind of science}. Champaign, IL: Wolfram Media. 	
	
	
	
	\bibitem{Zenil2020b} Zenil H. 2020 A Review of Methods for Estimating Algorithmic Complexity: Options, Challenges, and New Directions. \textit{Entropy} \textbf{22}, 612. (\href{http://doi.org/10.3390/e22060612}{doi:10.3390/e22060612})
	
	\bibitem{Zenil2020c} Zenil H, Kiani NA, Abrahão FS, Rueda-Toicen A, Zea AA, Tegnér J. 2020 Minimal Algorithmic Information Loss Methods for Dimension Reduction, Feature Selection and Network Sparsification. arXiv Preprints. Available at \href{https://arxiv.org/abs/1802.05843}{https://arxiv.org/abs/1802.05843}
	
	
	\bibitem{Zenil2020} Zenil H, Kiani N, Abrahão F, Tegnér J. 2020 Algorithmic Information Dynamics. \textit{Scholarpedia Journal} \textbf{15}, 53143. (\href{http://doi.org/10.4249/scholarpedia.53143}{doi:10.4249/scholarpedia.53143})
	
	

	
	
	
	
	
	\bibitem{Zenil2020a} Zenil H, Kiani NA, Marabita F, Deng Y, Elias S, Schmidt A, Ball G, Tegnér J. 2019 An Algorithmic Information Calculus for Causal Discovery and Reprogramming Systems. \textit{iScience} \textbf{19}, 1160–1172. (\href{http://doi.org/10.1016/j.isci.2019.07.043}{doi:10.1016/j.isci.2019.07.043})
	
	
	\bibitem{Zenil2017} Zenil H, Kiani NA, Tegnér J. 2017 Low-algorithmic-complexity entropy-deceiving graphs. \textit{Physical Review E} \textbf{96}, 012308. (\href{https://doi.org/10.1103/PhysRevE.96.012308}{doi:10.1103/PhysRevE.96.012308})
	
	
	
	\bibitem{Zenil2019a} Zenil H, Kiani NA, Tegnér J. 2019 The Thermodynamics of Network Coding, and an Algorithmic Refinement of the Principle of Maximum Entropy. \textit{Entropy} \textbf{21}, 560. (\href{http://doi.org/10.3390/e21060560}{doi:10.3390/e21060560})
	
		
	\bibitem{Zenil2019} Zenil H, Kiani NA, Zea AA, Tegnér J. 2019 Causal deconvolution by algorithmic generative models. \textit{Nature Machine Intelligence} \textbf{1}, 58–66. (\href{http://doi.org/10.1038/s42256-018-0005-0}{doi:10.1038/s42256-018-0005-0})
	
	
\end{thebibliography}
\bibliographystyle{plainnat}

\end{document}